\documentclass{article}
\pdfoutput=1
\usepackage{amsmath}
\usepackage{amssymb}
\usepackage{graphicx}
\usepackage{color}
\usepackage[usenames,dvipsnames,svgnames,table]{xcolor}
\usepackage{soul}
\usepackage{subfig}
\usepackage{ulem}
\usepackage{float}
\usepackage{multirow}

\pdfpageheight\paperheight
\pdfpagewidth\paperwidth

\usepackage{jcappub}

\definecolor{darkred}{RGB}{175,0,0}

\newcommand{\beq}{\begin{equation}}
\newcommand{\eeq}{\end{equation}}

\usepackage{array}
\def\ga{ \mathrel{\rlap{\raise 0.59ex
 \hbox{$>$}}{\lower 0.59ex \hbox{$\sim$}}}}

\subheader{\footnotesize
        \vspace*{-3.7em}
        \begin{flushright}
                CERN-TH-2018-223
        \end{flushright}
}
\begin{document}
\title{Probing $\Lambda$CDM cosmology with the Evolutionary Map of the Universe survey}
\author[a,b]{Jos\' e Luis Bernal}
\author[d,a,c]{Alvise Raccanelli}
\author[c]{Ely D. Kovetz}
\author[e]{David Parkinson}
\author[f,g]{Ray P. Norris}
\author[c]{George Danforth}
\author[c]{Courtney Schmitt}

\affiliation[a]{ICC, University of Barcelona, IEEC-UB, Mart\' i i Franqu\` es, 1, E08028
Barcelona, Spain}
\affiliation[b]{Dept. de F\' isica Qu\` antica i Astrof\' isica, Universitat de Barcelona, Mart\' i i Franqu\` es 1, E08028 Barcelona,
Spain}
\affiliation[c]{Department of Physics \& Astronomy, Johns Hopkins University, 3400 N. Charles St., Baltimore, MD 21218, USA}
\affiliation[d]{Theoretical Physics Department, CERN, 1 Esplanade des Particules, CH-1211 Geneva 23, Switzerland}
\affiliation[e]{Korea Astronomy and Space Science Institute, Yuseong-gu, Daedeok-daero 776, Daejeon 34055, Korea}
\affiliation[f]{Western Sydney University, Locked Bag 1797, Penrith South, NSW 1797, Australia}
\affiliation[g]{CSIRO Astronomy \& Space Science, PO Box 76, Epping, NSW 1710, Australia}

\emailAdd{joseluis.bernal@icc.ub.edu}
\emailAdd{alvise.raccanelli@cern.ch}
\emailAdd{ekovetz1@jhu.edu}
\emailAdd{davidparkinson@kasi.re.kr}
\emailAdd{raypnorris@gmail.com}

\abstract{The Evolutionary Map of the Universe (EMU) is an all-sky survey in radio-continuum which uses the Australian SKA Pathfinder (ASKAP). Using galaxy angular power spectrum and the integrated Sachs-Wolfe effect, we study the potential of EMU to constrain models beyond $\Lambda$CDM (i.e., local primordial non-Gaussianity, dynamical dark energy, spatial curvature and deviations from general relativity), for different design sensitivities. 
 We also include a multi-tracer analysis, distinguishing between star-forming galaxies and galaxies with an active galactic nucleus, to further improve EMU's potential. 
We find that EMU could measure the dark energy equation of state parameters around 35\% more precisely than existing constraints, and that the constraints on $f_{\rm NL}$ and modified gravity parameters will improve up to a factor $\sim2$ with respect to Planck and redshift space distortions measurements. 
With this work we demonstrate the promising potential of EMU to contribute to our understanding of the Universe.
}

\maketitle

\hypersetup{pageanchor=true}

\section{Introduction}\label{sec:Introduction}
Our current best description of the large-scale structure of the Universe relies on the standard cosmological model, $\Lambda$-Cold Dark Matter ($\Lambda$CDM), which posits that the energy density at present times is dominated by a cosmological constant and that the matter sector is composed mostly of dark matter. Although this model reproduces astonishingly well most observations \cite{Planck18_pars,Alam_bossdr12}, there are still some persistent tensions, especially on the Hubble constant between direct local measurements~\cite{RiessH0_2016,Riess18_GaiaDR2} and the Planck-inferred value assuming $\Lambda$CDM \cite{Planck18_pars}, which has been widely studied in the literature (see e.g., \cite{BernalH0,Bernal_baccus,Poulin_H0,DiValentino_interDE,DiValentino:2016hlg,Deramo_H0}). Moreover, there are also some theoretical issues within the model, such as the value  of the cosmological constant, the nature of dark matter and dark energy, an accurate description of inflation and the scale of validity of General Relativity. All this motivates the development and study of models beyond $\Lambda$CDM+GR. 

Up to this date, the strongest constraints on the parameters of $\Lambda$CDM have  come from Cosmic Microwave Background (CMB) observations. However, the Planck satellite has almost saturated the cosmic variance limit in the measurement of the CMB temperature  power spectrum. 
Moreover, low redshift observations are required in order to constrain models which extend $\Lambda$CDM to explain cosmic acceleration; in these cases, galaxy surveys are as of now the most powerful probe.
 The golden era of galaxy surveys is about to start, with some of the next generation experiments already observing or 
beginning in 2019. A huge experimental effort will provide game-changing galaxy catalogs, thanks to
which galaxy-survey cosmology will reach full maturity. 
Contrary to photometric or spectroscopic surveys, radio-continuum surveys average over all frequency data to have larger signal to noise for each individual source, which enables them to deeply scan large areas of the sky very quickly and detect faint sources at high redshifts. This allows the detection of large number of galaxies, but with only minimal redshift information.

Radio surveys have been used for cosmological studies in the past, mainly with NVSS~\cite{NVSS} (see e.g., ~\cite{Boughn_radio,Overzier_nvssclust,Boughn_isw,Nolta_isw,Smith_radiolensing,Raccanelli_radioisw,Ho_nvssisw,Afshordi_fNL,Xia_gammaray,Rubart_dipole,Giannantonio_isw13,Nusser_radioclust,Planck15_isw,Raccanelli_gw}). 
Studies of cosmological models and beyond-$\Lambda$CDM parameter constraints using next generation radio surveys were spearheaded in~\cite{Raccanelli_radio} and then followed by subsequent works such as e.g.,~\cite{Camera_radio,Raccanelli_iswfNL,Bertacca_udm,Jarvis_SKA,Camera_fNL,Ferramacho_radioPNG,Raccanelli2017_PNG,Karagiannis_PNG,Scelfo_progenitors,Ballardini_primordial,Alonso_ST}.
Forthcoming radio-continuum surveys have the unique ability to survey very large parts of the sky up to high-redshift, being therefore able to probe an unexplored part of the instrumental parameters space, not accessible to optical surveys for at least another decade. Thus, surveys like the Evolutionary Map of the Universe (EMU) and the Square Kilometer Array (SKA) continuum will be optimal for tests of non-Gaussianity, ultra-large scale effects and cosmic acceleration models, as we will see below.

 EMU \cite{EMU} is an all-sky radio-continuum survey using the Australian SKA Pathfinder (ASKAP) radio telescope \cite{Johnston_askap07, Johnston_askap08, Norris_ASKAP}. Although ASKAP was planned as a precursor of SKA to test and develop the needed technology, it is a powerful telescope in its own right. 
 In this work we aim to evaluate the potential of EMU as a cosmological survey, and in particular how powerful it can be in constraining extensions of $\Lambda$CDM. We pay special attention to primordial non-Gaussianity (PNG), since it manifests in the galaxy power spectrum on very large scales, accessible only by surveys like EMU, with large fractions of the sky observed. Constraining PNG is one of the few ways to observationally probe the epoch of inflation, and a precise measurement of its parameters might rule out a large fraction of inflationary models (e.g., slow-roll single-field inflation generally predicts small PNG, $f_{\rm NL}\ll 1$ for local PNG \cite{Bartolo_png, Komatsu_png, Wands_png}). Therefore, we explore if EMU will be able to detect deviations from Gaussian initial conditions in the local limit below $f_{\rm NL} \lesssim 1$~\cite{Alvarez_fNL}. Besides PNG, we also study a model of dark energy whose equation of state evolves with redshift; a model which does not fix the spatial curvature to be flat; and phenomenological, scale-independent  modifications of General Relativity (GR).

Given the lack of detailed redshift information, the main observable to be used with continuum radio surveys is the full shape of the angular galaxy power spectrum. Besides considering the whole sample altogether, we also take advantage of the expected broad distribution of sources in redshift and the potential of 
machine-learning redshift measurements (e.g.~\cite{Norris_MLredshift,Luken_MLredshift}) and clustering-based redshifts obtained by cross correlating EMU's sample with spectroscopic catalogues \cite{Menard_cbr,Rahman_cbr} (whose performance in cosmological surveys was estimated in~\cite{Kovetz_cbr}). These methods will enable to split the
sample into several redshift bins, and therefore we consider a second case where we use five redshift bins. 
 We leave the determination of the best strategy to the EMU redshift group. 
In addition to the auto and cross angular power spectra among all possible combination of redshift bins, we use the Integrated Sachs-Wolfe (ISW) effect by cross correlating the galaxies observed by EMU in each redsfhit bin with the CMB. 

This paper is organized as follows: in Section \ref{sec:EMU} we discuss the relevant  specifications of EMU and the estimated quantities required to compute the galaxy power spectrum, such as the source redshift distribution or galaxy bias; in Section \ref{sec:forecast} we explain the observables considered and the models studied, and describe the methodology used; results are discussed in Section \ref{sec:results}. We conclude in Section \ref{sec:discussion}. A detailed and comprehensive report of the results can be found in Appendix~\ref{App:tables}, and a comparison between using T-RECS~\cite{Bonaldi_TRECS} or $S^3$~\cite{Wilman_S3} can be read in Appendix~\ref{App:S3}. 

\section{EMU}\label{sec:EMU}
The main goal of EMU is to make a deep radio-continuum survey throughout the entire Southern sky reaching $\delta=+30^\circ$. This is roughly the same area covered by NVSS \cite{NVSS}, but approximately 45 times more sensitive  and with 4.5 times better angular resolution (it has a design sensitivity of 10 $\mu$Jy/beam rms and 10 arc-sec resolution). It is expected to observe around 70 million galaxies by the end of the survey. Thanks to these advantages, and especially to the large fraction of the sky scanned and the depth of
the survey, EMU (and future experiments of SKA) will be key for the study of galaxy       clustering at
 very large scales. 
 The corresponding catalog and the rest of radio data will be published once their quality has been assured.

We consider four different realizations of EMU, to compare the effectiveness of each:
\begin{itemize}
\item Design standard EMU: observing 30000 deg$^2$ of the sky (corresponding to a fraction of the whole sky $f_{\rm sky}=0.727$), with a sensitivity of 10 $\mu$Jy/beam rms.
\item Pessimistic EMU: the same areal coverage (30000 deg$^2$), with a sensitivity of 20 $\mu$Jy/beam rms.
\item EMU-early: an early stage of the survey with only 2000 deg$^2$ surveyed and 100 $\mu$Jy/beam rms.
\item SKA-2 {\it like} survey:  observing 30000 deg$^2$ of the sky with 1 $\mu$Jy/beam rms, to compare it with the design sensitivity of EMU.
\end{itemize}
 Like most of the studies regarding radio surveys, we adopt a 5$\sigma$ threshold for a source to be detected.
 
  We use the Tiered Radio Extragalactic Continuum Simulation (T-RECS) \cite{Bonaldi_TRECS} 
 to estimate the galaxy number density distribution as a function of redshift of both star forming galaxies (SFG) and galaxies with active galactic nucleus (AGN) and the corresponding galaxy and magnification biases. Throughout most of this paper, we work with the two population of galaxies together forming a single sample. However, we also estimate the gain of computing the auto power spectra of each population and their cross power spectrum using the multi-tracer technique \cite{Seljak_fNLmultitracer,McDonald_multitracer}. Although some studies distinguish between five different populations subdiving AGNs and SFGs (e.g., \cite{Ferramacho_radioPNG}), we prefer to be conservative and avoid that subdivision, since its robustness given the observational conditions is not very well quantified.

\begin{figure}
\centering
\includegraphics[width=0.49\linewidth]{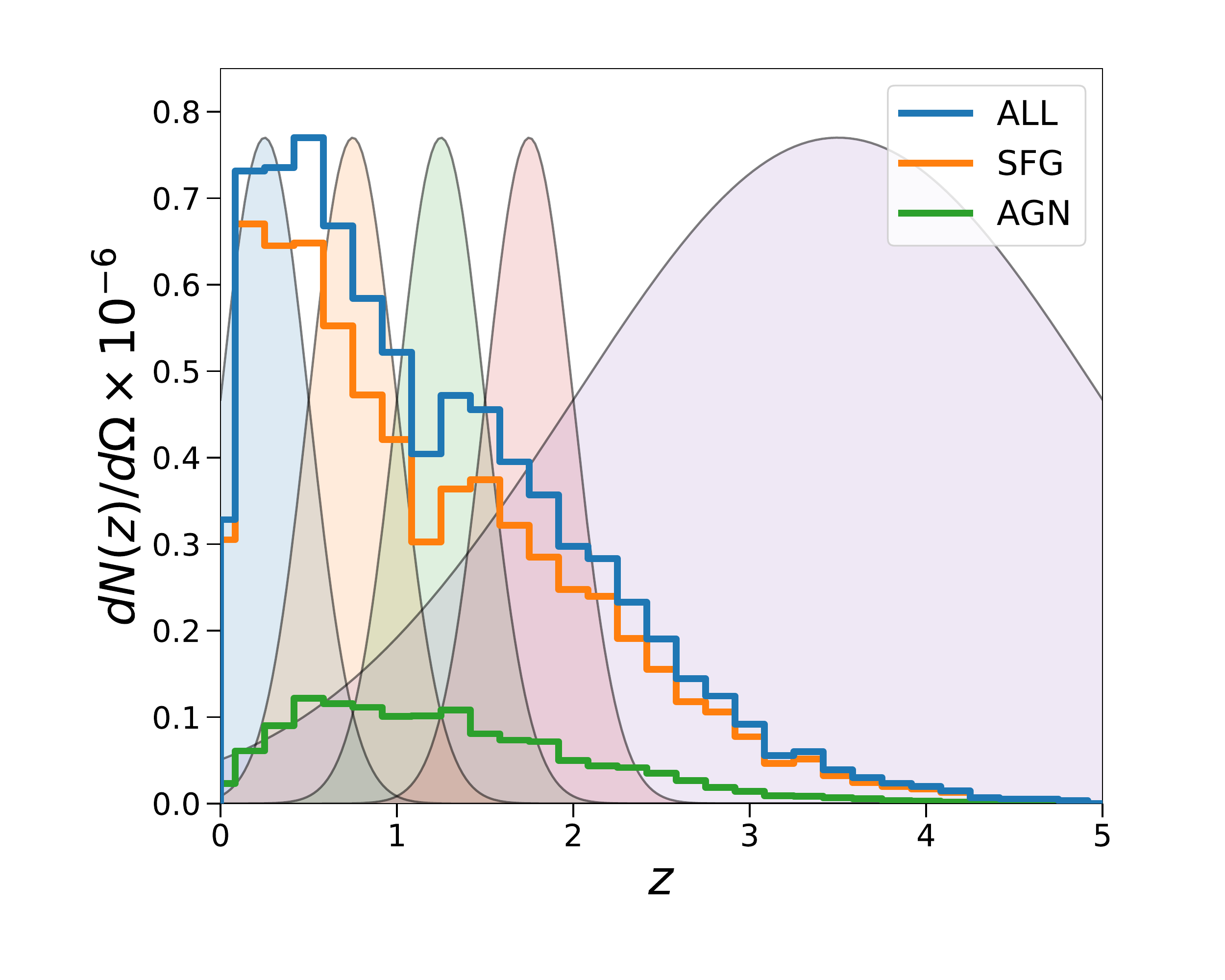}
\includegraphics[width=0.49\linewidth]{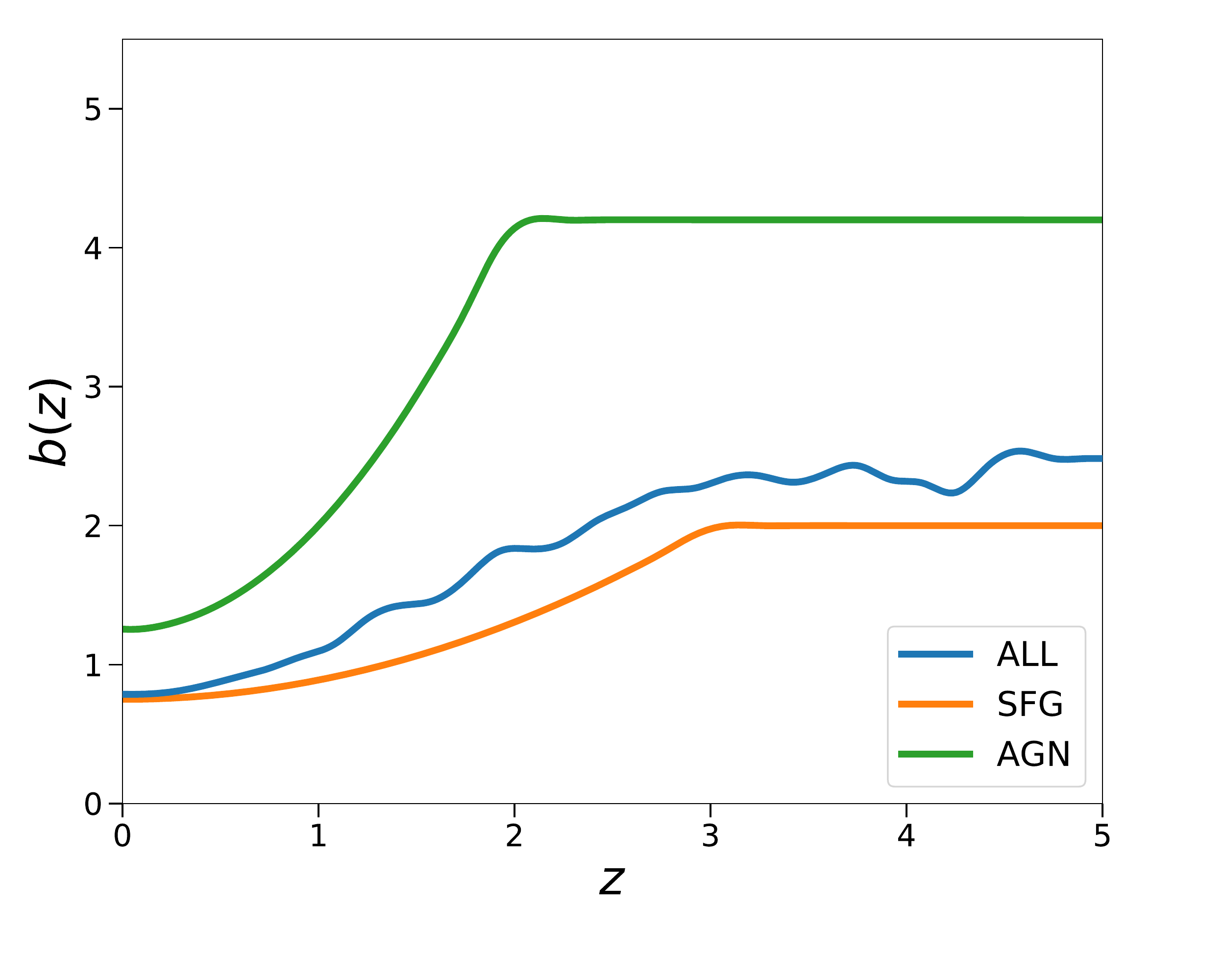}
\caption{ \textit{Left}: $dN/dzd\Omega$ for the galaxies observed assuming a rms flux/beam of 10$\mu$Jy, always requiring a 5-$\sigma$ detection. We also plot the Gaussian window functions of the five redshift bins considered to compute the cosmological observables.  \textit{Right} theoretical galaxy bias for SFGs and AGNs and the corresponding weighted total. We show the quantities related to the whole sample in blue, with SFGs in orange and with AGNs in green.}
\label{fig:specs}
\end{figure}

 As discussed above, the redshift determination of the galaxies observed by EMU, as a radio-continuum survey, will be very poor. This prevents the use of EMU to measure radial baryon acoustic oscillations or redshift space distortions, without precise external redshift information. Without any external data, it is not possible to determine the redshift of the sources and so bin the galaxy catalogue in redshift. Therefore, we consider a case with a single redshift bin covering the whole galaxy sample observed by EMU. Nonetheless, binning in redshift and using auto and cross correlations between the galaxies of all possible combinations of redshift bins improves significantly the performance of a galaxy survey. 
 We make a conservative choice and use five wide redshift bins with Gaussian window functions whose width is equal to the half width of the redshift bin, assuming that external data sets and the methodologies to infer the redshift are complete and mature enough to do so. Nonetheless, one should keep in mind that uncertainties in the galaxy properties as the galaxy bias degrades the quality of the inferred redshifts. We leave the study of the best redshift binning strategy of EMU's observed galaxies for future work. 
We show the total number density redshift distribution of galaxies, $dN/dzd\Omega$, for each galaxy population and for the complete sample at design sensitivity in the left panel of Figure \ref{fig:specs}, along with the corresponding Gaussian windows used for each redshift bin when computing the observables (see Section \ref{sec:probes}).  

In addition to modeling the galaxy redshift distribution, we need to model both the galaxy and magnification bias of our sample. We assume a different scale-independent galaxy bias model, which increases the bias monotonically with redshift, for each galaxy population in T-RECS. However, this trend becomes unphysical at high enough redshift, so here we follow \cite{Raccanelli_radiosurveys} and keep a constant bias above a cut-off redshift. 
 The true dependence of bias with redshfit at high-redshift, low-luminosity galaxies (where most of them are still undetected) is still unknown. On the other hand, the authors of \cite{Ferramacho_radioPNG} argue that the high-redshift cut-off is not physical. The choice of bias model is therefore a source of uncertainty, degenerate with the redshift distribution (see Section \ref{sec:probes} for further discussion). In order to compute the galaxy bias of the whole galaxy sample, we use the number-weighted average of both populations: $b(z) = [b_{\rm sfg}(z)N_{\rm sfg}(z) + b_{\rm agn}(z)N_{\rm agn}(z)]/N_{\rm all}(z)$.  We show the galaxy bias of SFGs, AGNs and the whole sample assuming 10 $\mu$Jy/beam rms sensitivity in the right panel of Figure \ref{fig:specs}. 

In order to estimate the magnification bias, we use the ``observed" magnitudes from T-RECS to get the slope of the cumulative number counts for galaxies brighter than the survey magnitude limit $m_{\rm lim}$ evaluated at $m_{\rm lim}$~\cite{Matsubara_lensing2000,Bartelmann_mag01,Liu_mag14}:
\begin{equation}
s(z) = \left. \frac{\partial}{\partial m} \left[{\rm log_{10}} n_{\rm cum} (m,z) \right] \right|_{m_{\rm lim}} \, ,
\label{eq:mag_bias}
\end{equation}
where $n_{\rm cum}$ is the cumulative number counts of galaxies as function of magnitude. As the number density evolves with redshift, this slope will also change, and so will the magnification bias. We compute the magnification bias for each population and the whole sample in the same way, using the corresponding $n_{\rm cum}(m,z)$ in each case. 

\section{Forecasts}\label{sec:forecast}
In this section we describe the cosmological observables included in our forecasts, introduce the models we investigate and review the Fisher matrix formalism used to predict the constraints. In all cases we assume a complete understanding of foregrounds and other sources of observational systematics, which would allow for a clean and precise measurement of galaxy clustering. We use \texttt{Multi\_CLASS}\footnote{\texttt{Multi\_CLASS} will be publicly available at \url{https://github.com/nbellomo?tab=projects} when the corresponding paper is published.}~\cite{Bellomo_Fisher}, a modification of the public Boltzmann code \texttt{CLASS}~\cite{Lesgourgues:2011re} which allows to compute the cross power spectra of two different populations, to obtain the theoretical observables.  

\subsection{Cosmological observables}\label{sec:probes}
We aim to estimate the potential of EMU to use galaxy clustering (measuring angular power spectra) to constrain models beyond $\Lambda$CDM. 
In addition, we also consider ISW measurements by cross correlating each redshift bin of our galaxy sample with the CMB. In all cases, we set a conservative cut on the minimum scale included, in order to avoid non linearities and limit the analysis to multipoles within the interval $\ell_{\rm min}\leq\ell\leq 200$. The largest scales considered, $\ell_{\rm min}$, are limited by the fraction of sky surveyed ($f_{\rm sky}$): $\ell_{\rm min}=\pi/(2f_{\rm sky})$. For both observables we also consider the case where SFGs and AGNs can be discriminated, allowing to use multi-tracer techniques.

The galaxy distribution, as we observe it, is affected by gravitational perturbations along the line of sight~\cite{Yoo_GR09}. Therefore, the total observed overdensity, $\Delta_{\rm obs}$, should include, in addition to the standard redshift-space distortions, large-scale and projection effects, namely lensing magnification, time-delays, ISW, doppler and gravitational potential effects~\cite{Yoo_GR10, Bonvin_obsLSS, Challinor_obsLSS, Jeong_GR12, Bertacca_xi3dGR, Raccanelli_GRCl}. 
 In this work we neglect contributions from gravitational potentials; the interested reader can see the complete model and corresponding signal and contributions from each term for different configurations in e.g., \cite{Raccanelli_GRint}. In our case, the observed galaxy overdensity at a position {\bfseries n} on the sky is computed as \cite{Bonvin_obsLSS,Challinor_obsLSS}:
\begin{equation}
\label{eq:delta_obs}
\Delta_{\rm obs} ({\bf n}, z) = \Delta_{\delta}({\bf n}, z) + \Delta_{\rm rsd}({\bf n}, z) + \Delta_{\rm \kappa}({\bf n}, z) + \Delta_{\rm Doppler}({\bf n}, z)  \, ,
\end{equation}
where $\Delta_\delta$ indicates the galaxy overdensity in the comoving gauge, $\Delta_{\rm rsd}$ accounts for peculiar velocity perturbations and redshift space distortions, $\Delta_\kappa$ contains the lensing convergence, and $\Delta_{\rm Doppler}$ refers to Doppler effects. The latter, although subdominant at small scales, is needed, since it is degenerate with $f_{\rm NL}$~\cite{Raccanelli_doppler}. 
Each of the terms in Equation~\eqref{eq:delta_obs} is given by:
\begin{equation}
\centering
\begin{split}
& \Delta_{\delta}({\bf n}, z)  = b(z)\delta_{\rm com}\left[r(z){\bf n},\tau(z)\right] \, , \\
&\Delta_{\rm rsd}({\bf n}, z)  = \frac{1}{{\cal H}(z)}\partial_r({\bf v}\cdot{\bf n}) \, , \\
&\Delta_{\rm \kappa}({\bf n}, z)  = \left[2-5s(z)\right]\kappa \, , \\
& \Delta_{\rm Doppler} = \left[ \frac{{\cal H}^\prime (z)}{{\cal H}^2(z)}+\frac{2-5s(z)}{r(z){\cal H} (z)} +5s(z) -f_{\rm evo}(z) \right] ({\bf v}\cdot{\bf n}) + \left[ 3{\cal H} (z)-f_{\rm evo}(z)   \right] \Delta^{-1}(\nabla \cdot{\bf v})\, ,
\end{split}
\label{eq:deltas}
\end{equation}
where $\delta_{\rm com}$ is the matter overdensity in the comoving gauge at a distance $r$ and proper time $\tau$, $\mathcal{H}$ is the conformal Hubble parameter, {\bf v} is the peculiar velocity, $\kappa$ is the lensing convergence and $f_{\rm evo}$ is the evolution bias. 

Using Equations \eqref{eq:delta_obs} and \eqref{eq:deltas}, we define $\Delta_\ell^{W_{\rm i}}$, the transfer function of the observed number counts at wavenumber $k$, in the redshift bin $i$ and using a window function $W_{\rm i}$ (Gaussian in our case) for each multipole $\ell$,  as in \cite{DiDio_Classgal}. We refer the interested reader to Appendix A of \cite{Raccanelli_GRCl} for details on the calculation of $\Delta_\ell^{W_i}$. 
With all these pieces, the observed angular power spectrum of galaxies in redshift bins $\{i,j\}$ is given by:
 \begin{equation}
  C_{\ell}^{\rm{ij}} = 4\pi \int \frac{dk}{k} \Delta_\ell^{W_{\rm i}}(k)\Delta_\ell^{W_{\rm j}}(k)P_0(k)\,,
  \label{eq:C_ell}
\end{equation}
where $P_0(k)$ is the primordial power spectrum. Note that when two non-overlapping bins are cross correlated, the distance is too large to have a significant contribution from intrinsic density clustering (i.e., coming from $\Delta_{\ell,\delta}^{W_{\rm i}}\Delta^{W_{\rm j}}_{\ell,\delta}$). However, we do observe a significant cross correlation due to lensing contributions~\cite{Raccanelli_radial}. Concretely, the dominant term is always $\Delta^{W_{\rm j}}_{\ell,\delta}\Delta_{\ell,\kappa}^{W_i}\propto b(z_{\rm j})s(z_{\rm i})$, which is the correlation between the magnification of background sources due to foreground lenses and the observed overdensities in the background. In some previous studies, magnification has been considered as a different signal than density perturbations (i.e., reporting the constraints from galaxy clustering and from magnification separately). However, this is not a realistic scenario, since these two contributions are difficult to disentangle, so we can only refer to the \textit{observed} galaxy clustering and model the signal properly. The effects of not including the lensing magnification contribution when modeling the signal in a Fisher forecast, even for cosmological parameters that are not affected by gravitational lensing, are studied in~\cite{Bellomo_Fisher}. 

When using the multi-tracer technique, we consider SFGs and AGNs as different galaxy populations, each with their corresponding redshift distribution, galaxy bias and magnification bias. Therefore, in this case we compute auto and cross correlations between different galaxy populations and redshift bins. Equation \eqref{eq:C_ell} can be modified so it also accounts for the two different galaxy populations. The transfer functions now become $\Delta_\ell^{{\rm X},W_i}$, where the ${\rm X}$ superscript refers to the type of galaxy, and they are computed as usual, but using the specifications corresponding for each subsample, as discussed in Section \ref{sec:EMU}. So, when using multiple tracers, we compute $C_\ell^{\rm X, i;Y, j}$.  

In addition to the angular power spectrum, we also use the ISW effect to measure the matter overdensity field. The ISW effect is the gravitational shift that a photon suffers as it passes through matter density fluctuations while the gravitational potential evolves. In an Einstein-de Sitter Universe, where the gravitational potential does not evolve, the blueshift and redshift of the photon falling and going out from a well cancel each other. Nonetheless, if the gravitational potential evolves due to e.g., dark energy or modifications of GR, the cancellation is not perfect, so there is a net change in the photon temperature which accumulates along the photon path. 

The ISW effect contributes to the CMB temperature fluctuations, but only on large scales, where the observations are limited by cosmic variance. However, it can also be detected in the cross correlation of the CMB anisotropies and the galaxy distribution. 
This correlation was detected for the first time almost simultaneously in several works using observations in radio from the NVSS survey~\cite{Nolta_isw,Boughn_isw,Giannantonio_isw08,Raccanelli_radioisw, Ho_nvssisw}; near infra-red from the 2-MASS survey~\cite{Afshordi_isw} and the APM survey~\cite{Fosalba_isw}; optical from SDSS~\cite{Scranton_isw}; and X-ray for HEAO-I satellite~\cite{Boughn_isw}.

 Equation \eqref{eq:C_ell} can be used to compute the cross correlations between the galaxy distribution and the CMB anisotropies for each redshift bin of our galaxy catalog, $C_\ell^{\rm iT}$, using $\Delta_\ell^{W_{\rm i}}(k)\Delta_\ell^{W_{\rm T}}(k)$ instead of $\Delta_\ell^{W_{\rm i}}(k)\Delta_\ell^{W_{\rm j}}(k)$, where $\Delta_\ell^{W_{\rm T}}(k)$ is the transfer function for CMB temperature anisotropies. Moreover, one can use multiple tracers and compute the cross correlations of the CMB with each galaxy population separately, in a similar fashion as for the angular galaxy power spectra.

\subsection{Models}\label{sec:models}
In this work we focus on popular models beyond the standard model of cosmology, which have one or two extra model parameters, since a low redshift, wide field survey as EMU will help significantly to constrain them thanks to the breaking of degeneracies existing in the CMB measurements. We focus on the following extensions of $\Lambda$CDM: a model with local PNG in the distribution of the initial conditions; a model with an evolving dark energy equation of state;  a model with scale-independent modifications of General Relativity; and a model where spatial curvature is not fixed to be flat.

\subsubsection{Primordial non-Gaussianity}
The ultra-large-scale modes of the matter power spectrum have remained outside the horizon since inflation. This is why they might preserve an imprint of primordial deviations from Gaussian initial conditions. Thanks to all-sky surveys, we can access  those ultra large scales and probe inflation models with low-redshift observations. We model PNG in the local limit, the easiest case to detect, introducing the parameter $f_{\rm NL}$\footnote{here we use the large scale structure convention ($f_{\rm NL}^{\rm LSS}\approx 1.3 f_{\rm NL}^{\rm CMB}$ \cite{Xia_PNG})}, defined as the amplitude of the local quadratic contribution of a single Gaussian random field $\phi$ to the Bardeen potential $\Phi$. 
We refer to this model as $\Lambda$CDM+$f_{\rm NL}$. 
 Other PNG models and the corresponding predicted constraints from a SKA-{\it like} survey can be found in e.g., \cite{Raccanelli2017_PNG}. In the limit considered here, the Bardeen potential is obtained as:
\begin{equation}
\Phi(x) = \phi(x) + f_{\rm NL}\left(\phi^2(x)-\langle\phi^2\rangle\right)\, .
\label{eq:bardeen}
\end{equation} 
The quadratic contribution in Equation \eqref{eq:bardeen} introduces skewness in the density probability distribution, which results in a modification of the number of massive objects. This can be modeled as a scale-dependent variation of the galaxy bias \cite{Matarrese_png00,Dalal_png07,Matarrese_png08,Desjacques_png}. If $b_{\rm G}$ is the Gaussian galaxy bias, the total galaxy bias is given by:
\begin{equation}
b(k,z) = b_{\rm G}(z) + \left[b_{\rm G}(z) - 1\right]f_{\rm NL}\delta_{\rm ec}\frac{3\Omega_m H_0^2}{c^2k^2T(k)D(z)}\, ,
\label{eq:bias_PNG}
\end{equation}
where $\delta_{\rm ec}=1.68$ is the critical value of the matter overdensity for spherical collapse, $\Omega_m$ is the matter density parameter at $z=0$, $D(z)$ is the linear growth factor (normalized to 1 at $z=0$) and $T(k)$ is the matter transfer function (which is 1 on large scales). Thus, the deviation from the Gaussian galaxy bias at small $k$ is proportional to $f_{\rm NL}k^{-2}$, hence it contributes significantly only on large scales. 

\subsubsection{Dynamical dark energy}
Dark energy can be modeled with a scalar field, instead of a cosmological constant as in $\Lambda$CDM. In that case, the equation of state of  dark energy, $w$, may be different than $-1$ and also vary with redshift. The energy density of  dark energy is then no longer constant and is given by
\begin{equation}
\rho_{\rm DE}(a) = \rho_{{\rm DE},0}\exp\left[ -3\int_1^a \frac{1+w(a')}{a'}da' \right],
\end{equation}
where $\rho_{{\rm DE},0}$ is the density of dark energy today. 
We use the CPL parameterization \cite{CPL_w0wa, Linder_w0wa} to model $w(a)$  as:
\begin{equation}
w(a) = w_0+(1-a)w_a,
\label{eq:CPL}
\end{equation}
where $a$ is the scale factor. Therefore, we call this model $(w_0w_a)$CDM. 

\subsubsection{Modified gravity}
Although cosmic acceleration is normally modeled using dark energy, it can also be explained in theory with modifications to gravity. Moreover, GR might be a local approximation, and has only been tested precisely on scales ranging from millimeters to solar-system scales, with a compelling test at horizon scales yet to be done. 
 This is what has motivated the theoretical development of alternative theories to GR, adding degrees of freedom.  
There is a huge variety of modified gravity theories, although a large fraction of them are ruled out after the recent measurement of the neutron-star merger and gravitational-wave counterpart (see e.g., \cite{Ezquiaga_GWreview} for an updated  review). Moreover, consistency tests of GR using current data do not favour modifications (see e.g., \cite{Bernal_ParamSplit}). Nonetheless, GR might not be the correct description of gravity at the ultra-large scales that will be surveyed by EMU. 
 In order to model deviations from GR in a general way, we follow an effective description of the relation between the metric potentials and their relation with the energy density \cite{Amendola_MG,Zhao_MG}:
\begin{equation}
-2k^2\Psi=8\pi G_Na^2\rho \delta_{\rm com}\mu(a,k)\,, \qquad\qquad \frac{\Phi}{\Psi} = \gamma(a,k)\,,
\label{eq:MG}
\end{equation}
where $\mu=\gamma=1$ are the limiting values corresponding to GR,  where $\rho$ (the total energy density) and $\delta_{\rm com}$ are evaluated at $a$. We only consider scale independent modifications of GR, but there are several possible parameterizations of $\mu(z,k)$ and $\gamma(z,k)$~\cite{Baker_mgclass,Planck18_pars}. We assume that deviations from GR are only significant at low redshifts, so we model them as being proportional to the dark energy density parameter, $\Omega_\Lambda$
\begin{equation}
\mu(a,k) = 1+\mu_0\frac{\Omega_\Lambda(a)}{\Omega_{\Lambda,0}}, \qquad \gamma(a,k) = 1+\gamma_0\frac{\Omega_\Lambda(a)}{\Omega_{\Lambda,0}}, 
\label{eq:mugamma}
\end{equation}
where $\Omega_{\Lambda,0}$ is the dark energy density parameter today. 
Combining Equations \eqref{eq:MG} and \eqref{eq:mugamma}, we obtain that with this parameterization  $\Lambda$CDM with GR corresponds to $\mu_0 = \gamma_0 = 0$. 
In this work, this model is referred to as $\Lambda$CDM+$\mu_0$+$\gamma_0$. We follow \texttt{MGCLASS}\footnote{\url{https://gitlab.com/philbull/mgclass}}~\cite{Baker_mgclass} and modify  \texttt{Multi\_CLASS} to include this parameterization of modified gravity in our computations.  

Finally, we also consider a model in which the curvature of the spatial sector of the Universe is constant, but not fixed to be zero. We denote this model $\Lambda$CDM+$\Omega_k$.

\subsection{Fisher matrix formalism}\label{sec:fisher}
In order to forecast the constraining power of EMU for the models discussed above, we use the Fisher matrix analysis \cite{Fisher:1935,Tegmark_fisher97}. Accounting for SFGs and AGNs as different tracers of the dark matter field, we define $\tilde{C}_\ell^{\rm X,i;Y,j}$ as the angular power spectrum plus the shot noise as:
\begin{equation}
\tilde{C}_\ell^{\rm X,i; Y,j} = C_\ell^{\rm X,i;Y,j}+\frac{\delta^K_{\rm ij}\delta^K_{\rm XY}}{dN(z_{\rm i})/d\Omega},
\label{eq:C_ell_tilde}
\end{equation}
where $\delta^K$ is the Kronecker delta and $dN(z_{\rm i})/d\Omega$ denotes the average number of sources per steradian in the $i$-th redshift bin. If we assume a Gaussian likelihood, it is possible to define a covariance matrix for each multipole $\mathcal{C}_\ell$ built by blocks, so if each block is indexed by $\{X, Y\}$, $\left(\mathcal{C}_\ell\right)_{\rm X,Y} = \tilde{C}_\ell^{\rm XY}$. Then, each block is built as $\left(\tilde{C}_\ell^{\rm XY}\right)_{i,j}=\tilde{C}_\ell^{\rm X,i; Y,j} $. 
 In this way, the Fisher matrix element corresponding to the parameters $\theta_\alpha$ and $\theta_\beta$ for the galaxy angular power spectrum is given by:
\begin{equation}
F_{\alpha\beta}^{\rm gg} = \left\langle \frac{\partial^2\log\mathcal{L}^{\rm gg}}{\partial\theta_\alpha\partial\theta_\beta}\right\rangle = f_\mathrm{sky}\sum_\ell  \frac{2\ell+1}{2} \mathrm{Tr}\left[\frac{\partial \mathcal{C}_\ell}{\partial\theta_\alpha}\mathcal{C}^{-1}_\ell\frac{\partial \mathcal{C}_\ell}{\partial\theta_\beta}\mathcal{C}^{-1}_\ell\right].
\label{eq:fisher_gg_tr}
\end{equation}

Using Equation \eqref{eq:fisher_gg_tr} ensures that the full covariance of all galaxy angular power spectra considered is accounted for properly. If the covariance between different power spectra was neglected, the results of the Fisher analysis may change dramatically, as shown in \cite{Bellomo_Fisher}.

We consider the ISW effect as an independent cosmological probe. 
  Moreover, the ISW computed for each redshift bin and each of the galaxy population is independent from the other (when considered separately). In this case, we also assume a Gaussian likelihood and then the Fisher matrix is obtained as:
\begin{equation}
F_{\alpha\beta}^{\rm ISW} = \sum_{\ell\ell^\prime, {\rm X, i}} \frac{\partial C_\ell^{\rm X,i;T}}{\partial\theta_\alpha}\frac{\partial C_{\ell^\prime}^{\rm X,i;T}}{\partial\theta_\beta}\delta^K_{\ell\ell^\prime}\sigma^{-2}_{C_\ell^{\rm X,i;T}}\, , \qquad \sigma_{C_\ell^{\rm X,i;T}} = \sqrt{\frac{\left(C_\ell^{\rm X,i;T}\right)^2+\tilde{C}_\ell^{\rm X,i; X,i}C_\ell^{\rm TT}}{(2\ell+1)f_{\rm sky}}}\,,
\label{eq:fisher_isw}
\end{equation}
where $C_\ell^{\rm TT}$ is the CMB temperature angular power spectrum, and we neglect the error of the CMB measurement beyond cosmic variance, since it is much smaller than the shot noise of the galaxy power spectra.

 Assuming a Gaussian likelihood for the $C_\ell$ is a good approximation for large $\ell$, since the central limit theorem can be applied due to the large number of modes. However, on large scales, where the number of modes is limited, the true likelihood is better approximated by a lognormal likelihood (see e.g., \cite{Verde_wmap03}). Nonetheless, Fisher matrix analysis assumes Gaussianity; one should use the extension proposed in \cite{Sellentin_dali} to account for non Gaussian likelihood. In any case, we do not expect large changes, since Fisher forecasts overestimate errors in this case (compared with the lognormal likelihood) and we are considering a complete understanding and removal of systematics. As systematics affect more the observations on large scales, these effects cancel each other qualitatively, justifying  the Gaussian approximation for the likelihood in this case. We leave the exploration of the effect of a non Gaussian likelihood to future work.

\begin{table}[]
\small
\centering
\begin{tabular}{|c|c|c|}
\hline
Parameter              & Meaning                                                                                                                                              & Equation                                                        \\ \hline
$f_{\rm NL}$           & Non Gaussian parameter in the local limit                                                                                                            & Eq. \eqref{eq:bardeen}                           \\ \hline
$w_0$                  & Equation of state of the dark energy fluid at redshift 0                                                                                             & Eq. \eqref{eq:CPL}                               \\ \hline
$w_a$                  & \begin{tabular}[c]{@{}c@{}}Amplitude of the time varying contribution \\ to the equation of state of the dark energy fluid\end{tabular}               & Eq. \eqref{eq:CPL}                               \\ \hline
$\Omega_k$             & Spatial curvature energy density parameter                                                                                                           & -                                                               \\ \hline
$\mu_0$                & \begin{tabular}[c]{@{}c@{}}Amplitude of the time varying ($\propto \Omega_\Lambda(a)$) deviation \\ from $\mu=1$ in modified gravity\end{tabular}    & Eq. \eqref{eq:mugamma}                           \\ \hline
$\gamma_0$             & \begin{tabular}[c]{@{}c@{}}Amplitude of the time varying ($\propto \Omega_\Lambda(a)$) deviation\\  from $\gamma=1$ in modified gravity\end{tabular} & Eq. \eqref{eq:mugamma}                           \\ \hline
$\Delta b_{\rm all}^X$ & \begin{tabular}[c]{@{}c@{}} Uncertainty on the galaxy bias of tracer $X$ \\ (or all the sample when needed).   \end{tabular}                                                                        & $b_{\rm true}^X(z) = b^X(z)+\Delta b^X_{\rm all}$               \\ \hline
$\Delta b_{\rm i}^X$   & \begin{tabular}[c]{@{}c@{}}Uncertainty on the galaxy bias\\ of tracer $X$ (or all the sample when needed) in the redshift bin i.            \end{tabular}                                                                      & $b_{\rm true}^X(z_{\rm i}) = b^X(z_{\rm i})+\Delta b^X_{\rm i}$ \\ \hline
$\Delta s_{\rm i}^X$   & \begin{tabular}[c]{@{}c@{}}Uncertainty on the magnification bias \\ of tracer $X$ (or all the sample when needed) in the redshift bin i.                                                                          \end{tabular} & $s_{\rm true}^X(z_{\rm i}) = bsX(z_{\rm i})+\Delta s^X_{\rm i}$ \\ \hline
\end{tabular}
\caption{Reference and meaning of each of the symbols used to denote the parameters included in the Fisher Matrix analysis along with the five standard cosmological parameters needed to compute the galaxy clustering.}
\label{tab:params}
\end{table}

As discussed in Section \ref{sec:models}, we will explore motivated models beyond $\Lambda$CDM, focusing on the extra cosmological parameters. Nonetheless, we also vary the five relevant parameters of $\Lambda$CDM for the galaxy clustering. 
In addition, we consider different levels of knowledge about the properties of each galaxy population, regarding galaxy and magnification bias. First, we assume a complete knowledge of $b^{\rm X}(z)$ and $s^{\rm X}(z)$.
Secondly we include a single parameter, $\Delta b_{\rm all}^X$, to model our ignorance with respect to the galaxy bias at all redshifts.
Third, we repeat the same strategy, but having an independent $\Delta b^X_{\rm i}$ for each redshift bin.
Finally, we also add independent $\Delta s^X_{\rm i}$ in a similar fashion to model our ignorance with respect to the magnification bias. In summary, if $\vec{\beta}$ denotes the parameters included to model our ignorance about the galaxy populations properties, we will have four cases: $\vec{\beta}=\varnothing$; $\vec{\beta}=\lbrace\Delta b_{\rm all}^X\rbrace$; $\vec{\beta}=\lbrace\Delta b_{\rm i}^X\rbrace$; and $\vec{\beta}=\lbrace \Delta b_{\rm i}^X, \Delta s_{\rm i}^X\rbrace$. This range of possibilities is a fair estimate, since the galaxy and magnification biases cannot be perfectly measured, but we will not be completely ignorant about them either, which is the case corresponding to the marginalization over  $\lbrace \Delta b_{\rm i}^X, \Delta s_{\rm i}^X\rbrace$. The most realistic scenario will be somewhere in between. All the definitions and relevant equations or corresponding references regarding the extra cosmological parameters and the nuisance galaxy bias and magnification bias parameters for each of the cases discussed above can be found in Table \ref{tab:params}. 

 Therefore, if a given model has $\vec{\Upsilon}$ extra cosmological parameters with respect to $\Lambda$CDM, we will consider the next set of parameters for the Fisher matrix:
\begin{equation}
\label{eq:params}
\vec{\theta} = \left\lbrace \Omega_{\rm b}h^2,\Omega_{\rm cdm}h^2, h, n_s, \log \left(10^{10}A_s\right),\vec{\Upsilon},\vec{\beta}\right\rbrace\, ,
\end{equation}
where $\Omega_{\rm b}h^2$ and $\Omega_{\rm cdm}h^2$ are the baryon and cold-dark-matter physical densities, respectively, $h$ is $H_0/(100\, {\rm km/s/Mpc})$ (with $H_0$ being the Hubble constant), and $n_s$, $A_s$ are the spectral index and the amplitude of the primordial power spectrum of scalar modes, respectively. 
We assume a $\Lambda$CDM model as our fiducial cosmology, setting the cosmological parameter fiducial values to the best fit of the analysis of the temperature, polarization and lensing power spectra~\cite{Planck18_pars} 
and BAO~\cite{Alam_bossdr12,Beutler11,Ross15} assuming the base model\footnote{The results of this analysis are denoted as \texttt{base\_plikHM\_TTTEEE\_lowl\_lowE\_lensing\_post\_BAO} at  the public Planck repository \href{http://pla.esac.esa.int/pla/}{http://pla.esac.esa.int/pla/}, where also the public MCMC can be found.}: $\Omega_{\rm b}h^2 = 0.022447$, $\Omega_{\rm cdm}h^2=0.11928$, $h=0.67702$, $\log (10^{10}A_s)=3.048$ , and $n_s = 0.96824$.

\section{Results}\label{sec:results}
In this Section, we discuss the results of the Fisher matrix analysis; a detailed report of the results can be found in Tables~\ref{tab:constr_fLN},~\ref{tab:constr_w0wa},~\ref{tab:constr_omegak},~\ref{tab:constr_MG} and~\ref{tab:constr_multi} in  Appendix~\ref{App:tables}. 
 As stated above, we forecast constraints from EMU considering  four different realizations regarding the rms flux per beam (using always a threshold of five times the rms flux per beam to claim a detection) and the fraction of sky covered. We use two cosmological probes: the angular power spectrum of the observed galaxies 
 (referred to as galaxy clustering, GC, in the plots and tables of this section) and the ISW effect (see Section \ref{sec:probes} for more details).  We estimate the constraints for all realizations of EMU considered using all galaxies as a single tracer, but we also consider two different tracers for the design sensitivity case of EMU. We expect the gains obtained in this case thanks to the multi-tracer technique to be equivalent in the other realizations considered in this work. 
 
 In order to estimate the combined constraints with current observations, we add priors from the temperature, polarization and lensing CMB power spectra from Planck~\cite{Planck18_pars},  combined with BAO observations from spectroscopic galaxy surveys~\cite{Alam_bossdr12,Ross15,Beutler11} (except for $\Lambda$CDM+$\mu_0$+$\gamma_0$, for which we use the constraints from Planck+BAO+RSD, as done in~\cite{Planck18_pars}). 
 We take the priors from current observations from the publicly available Monte-Carlo Markov Chains (MCMCs) of Planck collaboration 
 when possible. 
 However there are no public MCMCs for any modified gravity model or $\Lambda$CDM+$f_{\rm NL}$. For these cases, we make use of the marginalized 68\% confidence level constraints on each of the extra parameters reported in \cite{Planck18_pars} and \cite{Planck15_PNG}, respectively, to the $\Lambda$CDM parameter covariance matrix, only in the corresponding diagonal element. The main limitation of this procedure is the lack of information of the degeneracies between the extra parameters and the $\Lambda$CDM parameters (and also  between the extra parameters in the case of the modified gravity model). However, we do not expect that our forecast depends significantly on these degeneracies. The exploration of the effect of these degeneracies is left for future work.

\begin{figure}
\centering
\includegraphics[width=\linewidth]{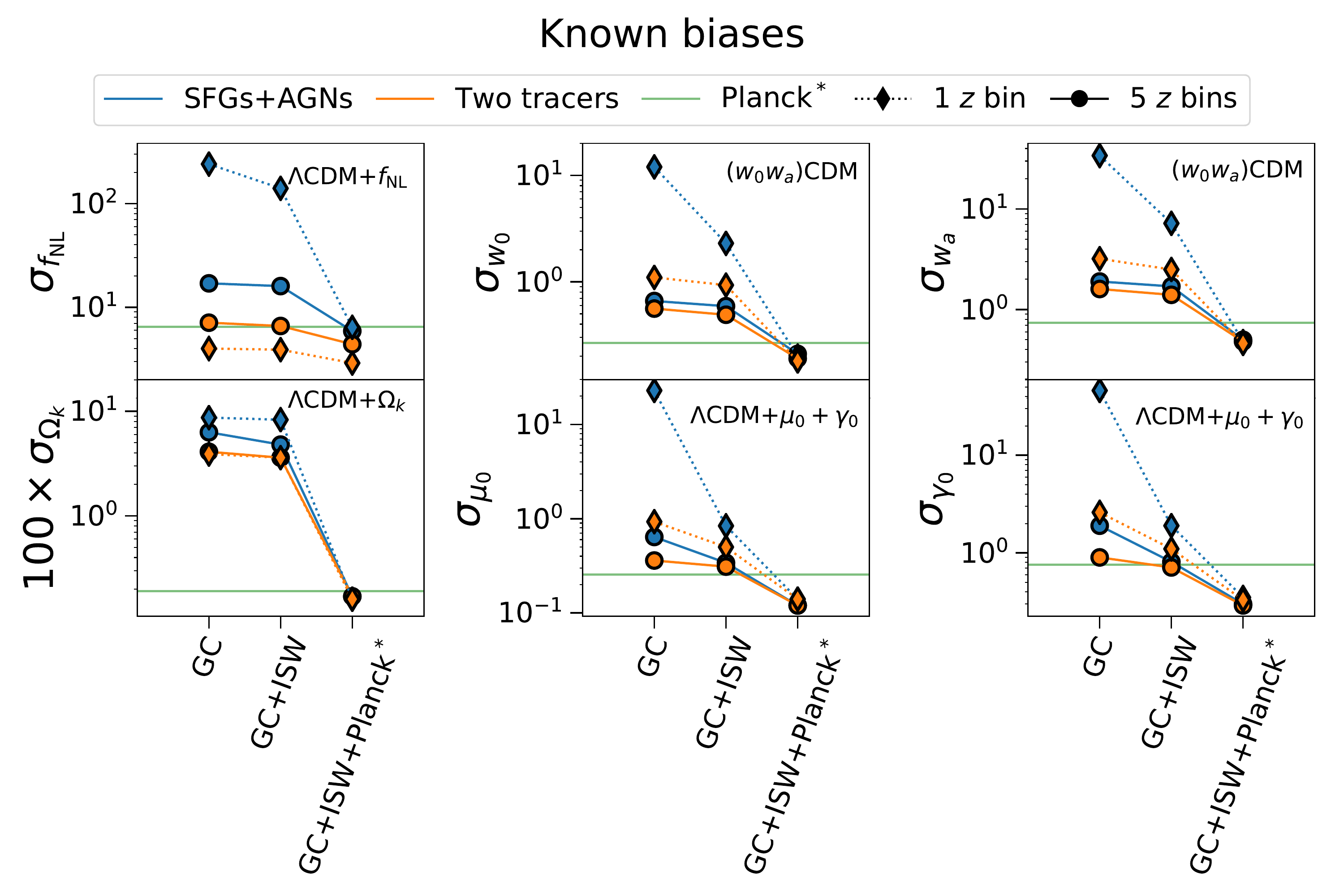}
\caption{Marginalized 68$\%$ confidence level forecast constraints around the $\Lambda$CDM limit values on each of the extra cosmological parameters of the models discussed in Section~\ref{sec:models}, for different data combinations, considering a realization of EMU at design sensitivity, without marginalizing over galaxy and magnification bias uncertainty parameters. We show results assuming one (dotted lines with diamonds) or five redshift bins (solid lines with circles) and using all the galaxies as a single tracer (blue lines) or discriminating between SFGs and AGNs (orange lines). We also show current constraints in green. ``+Planck$^*$'' means Planck+BAO in all cases but in the $\Lambda$CDM+$\mu_0$+$\gamma_0$ constraints, where it means Planck+BAO+RSD. Note the change of scale in the vertical axis in all cases.}
\label{fig:constraints}
\end{figure}

\begin{figure}
\centering
\includegraphics[width=\linewidth]{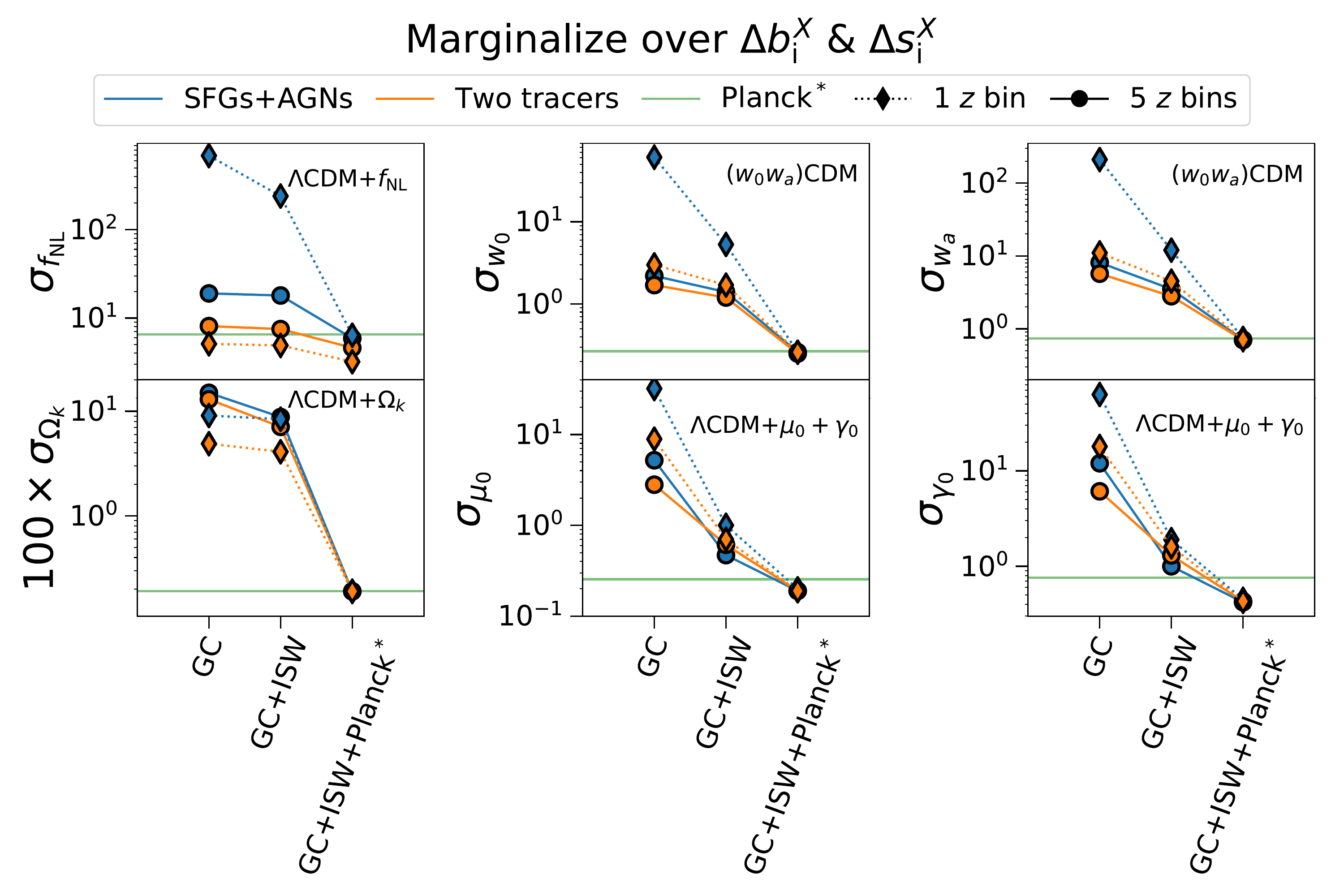}
\caption{Same as Figure \ref{fig:constraints} but marginalizing also over all the galaxy and magnification biases uncertainty parameters: $\Delta b_{\rm i}$ and $\Delta s_{\rm i}$.}
\label{fig:constraints_nbins}
\end{figure}

We show  in Figure~\ref{fig:constraints} the marginalized forecast constraints for the extra parameters of each extension of $\Lambda$CDM discussed in Section \ref{sec:models}, assuming both the galaxy and magnification bias are completely understood. On the other hand, we also show  in Figure \ref{fig:constraints_nbins} similar constraints marginalizing over $\Delta b_{\rm i}^X$ and $\Delta s_{\rm i}^X$. The most realistic scenario lies between these two cases, since some knowledge of the galaxy and magnification biases of the observed sources is expected by the time EMU is finished. We consider EMU at design sensitivity (i.e. 10 $\mu$Jy of rms flux per beam and 30000 deg$^2$ of sky scanned) for different data combinations: angular galaxy power spectra alone, adding ISW and adding ISW and priors of current constraints. In each case, we report results taking the whole sample in a single redshift bin (dotted lines with diamonds) and splitting the catalogue in five redshift bins (solid lines with circles); and also using SFGs and AGNs as the same tracer (blue lines) and as different tracers (orange lines). We moreover show current constraints with a green line in order to compare it with EMU's forecasts. In all cases (no matter the redshift binning, the number of tracers or the sensitivity of the survey), EMU will not be competitive with current constraints on $\Omega_k$ from Planck+BAO, so we will not discuss these results in this Section. Results for constraints on curvature are shown in Appendix~\ref{App:tables}. 

\begin{figure*}[htb]
\includegraphics[width=0.9\linewidth]{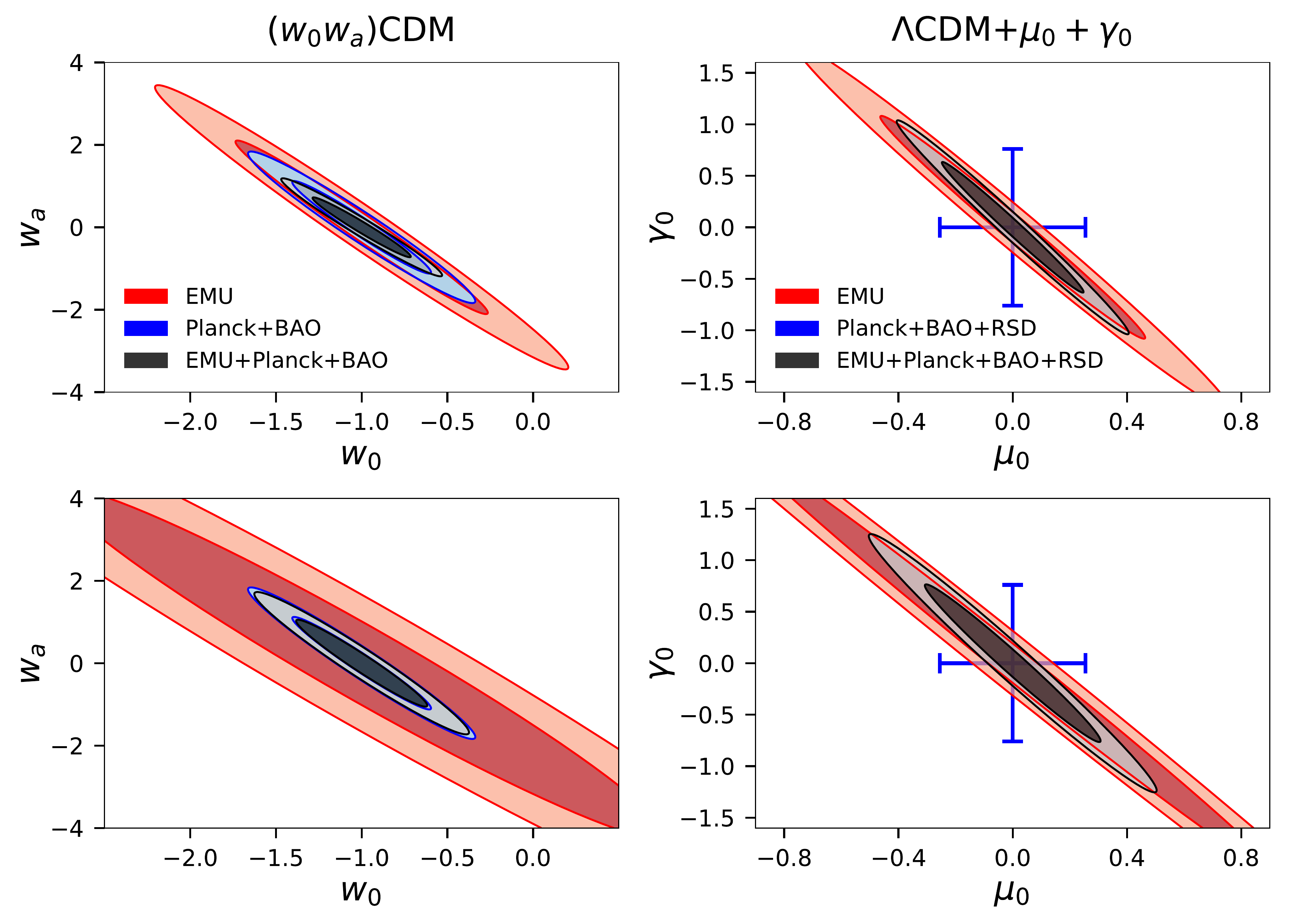}
\caption{68$\%$ and 95$\%$ confidence level forecast constraints around the $\Lambda$CDM limit values on $w_0$ and $w_a$ in a ($w_0w_a$)CDM cosmology (left panels) and on $\mu_0$ and $\gamma_0$ for $\Lambda$CDM+$\mu_0$+$\gamma_0$ (right panels).
EMU at design sensitivity are in red, Planck+BAO in blue, and their combination in black.  Upper panels assume total knowledge of galaxy and magnification bias, bottom panels marginalize over them. 
We only show one-dimensional marginalized 68\% confidence level constraints from Planck+BAO+RSD in the right panels because there is no public result on the correlation between $\mu_0$ and $\gamma_0$, nor public MCMC for this model.}
\label{fig:contours}
\end{figure*}

Assuming that the discrimination between SFGs and AGNs is not good enough to perform a multi-tracer analysis and that the redshift inference using external data sets is not reliable enough in order to split the sample in different redshift bins, EMU's constraints will not be competitive on their own with current observations. 

\begin{table}[]
\centering
\begin{tabular}{|c|c|c|c|c|c|}
\hline
\multicolumn{6}{|c|}{10 $\mu$Jy rms/beam, 1 Tracer, 1 Redshift bin}                                           \\ \hline
                  & $f_{\rm NL}$ & $w_0$       & $w_a$       & $\mu_0$     & $\gamma_0$  \\ \hline
GC+ISW            & 140 (240)    & 2.3 (5.3)   & 7.2 (12)    & 0.84 (1.0)  & 1.9 (1.9)   \\ \hline
GC+ISW+Planck$^*$ & 6.4 (6.4)    & 0.20 (0.26) & 0.48 (0.74) & 0.14 (0.20) & 0.35 (0.45) \\ \hline
\end{tabular}
\caption{Marginalized 68\% confidence level predicted constraints for the extra parameters of the models $\Lambda$CDM+$f_{\rm NL}$, $\left( w_0w_a\right)$CDM and $\Lambda$CDM+$\mu_0+\gamma_0$, from radio-continuum measurements with EMU, assuming design sensitivity and using one redshift bin and tracer. 
 We report constraints assuming total knowledge about galaxy and magnification biases and marginalizing over the corresponding parameters $\Delta b_{\rm all}$ and $\Delta s_{\rm all}$, in parentheses. 
We show results including galaxy power spectra and ISW effect (GC+ISW), and also including Planck+BAO priors (Planck+BAO+RSD in the case of $\Lambda$CDM+$\mu_0+\gamma_0$) denoted as Planck$^*$.}
\label{tab:1z1trac}
\end{table}

Combining the angular galaxy power spectrum and the ISW effect and assuming a total knowledge of the galaxy and magnification bias (marginalizing over $\Delta b_{\rm all}$ and $\Delta s_{\rm all}$, we obtain the following forecasts: $\sigma (f_{\rm NL}) = 140$ (240), $\sigma (w_0) = 2.3$ (5.3), $\sigma (w_a) = 7.2$ (12), $\sigma (\mu_0) = 0.84$ (1.0) and $\sigma(\gamma_0) = 1.9$ (1.9), assuming the corresponding model. 

We report the predicted constraints on the extra parameters in Table \ref{tab:1z1trac} assuming EMU observations at design sensitivity and using only one tracer and one redshift bin. We report results assuming total knowledge about the galaxy and magnification bias and also assuming total ignorance and marginalizing over the corresponding parameters, in parenthesis. In this configuration, EMU will not be competitive on its own. However, when combined with current constraints from CMB, BAO and RSD observations, EMU's measurements will improve current constraints on the modified gravity model a 45\% (22\%) for $\mu_0$ and a 54\% (41\%) for $\gamma_0$. In this case, they will also improve mildly the constraints on $\left( w_0w_a\right)$CDM, only if the galaxy and magnification biases are completely known:   25\% for $w_0$ and  35\% for $w_a$.

 Nonetheless, the improvement of redshift inference methodologies with external data sets and the advent of new galaxy catalogs with better sensitivity will allow to determine the redshifts with higher precision. Thanks to this, splitting EMU's catalogue in five redshift bins and using SFGs and AGNs as two different tracers is expected to be feasible. We report the resulting predicted constraints for this configuration in Table~\ref{tab:5z2trac}, in a similar fashion as before. The strongest constraints set by the combination of EMU observations with current data will be a factor 2.3 (2.0) for $f_{\rm NL}$,  33\% (equal) and  37\% (7\%) stronger for $w_0$ and $w_a$, respectively, and a factor 2.1 (1.3) and 2.3 (1.8) smaller for $\mu_0$ and $\gamma_0$, respectively, than current observations when the galaxy and magnification bias are assumed to be known (marginalizing over $\Delta b_{\rm i}^X$ and $\Delta s_{\rm all}^X$).

\begin{table}[]
\centering
\begin{tabular}{|c|c|c|c|c|c|}
\hline
\multicolumn{6}{|c|}{10 $\mu$Jy rms/beam, 2 Traces, 5 Redshift bins$^a$}                     \\ \hline
                  & $f_{\rm NL}$ & $w_0$       & $w_a$       & $\mu_0$     & $\gamma_0$  \\ \hline
GC+ISW            & 3.9 (4.9)    & 0.49 (1.2)  & 1.4 (2.8)   & 0.31 (0.61) & 0.71 (1.3)  \\ \hline
GC+ISW+Planck$^*$ & 2.9 (3.2)    & 0.19 (0.25) & 0.48 (0.70) & 0.12 (0.19) & 0.29 (0.43) \\ \hline
\end{tabular}
\caption{Same as Table~\ref{tab:1z1trac} but using  five redshift bins and two tracers.\\
$^a$: In the case of $\Lambda$CDM+$f_{\rm NL}$ we report the result using a single redshift bin, since it is stronger in this case. }
\label{tab:5z2trac}
\end{table}

As expected, using two tracers improves EMU's performance, especially for $f_{\rm NL}$, where EMU alone constraints are stronger than current bounds. 
This is not surprising, since PNG imprints appear on the largest scales, those more affected by the cosmic variance, which is partially overcome by the multi-tracer technique. What it is surprising is that using SFGs and AGNs as different tracers the constraints on $f_{\rm NL}$ are better using only one redshift bin. This is due to the fact that PNG imprints are very sensitive to $b(z)$. In Figure~\ref{fig:specs} it is shown that AGN galaxy bias is much larger than SFG galaxy bias. However, EMU will observe many more SFGs than AGNs, and it will not detect enough AGNs in the high-redshift bins. Therefore, for the AGN power spectra at high redshifts, where the impact of PNG is larger, the shot noise is also larger and the final sensitivity of EMU to $f_{\rm NL}$ decreases with respect to having only one bin. 
 Then, the strongest constraints on $f_{\rm NL}$ found in this work correspond to having two tracers with only one redshift bin. Something similar happens for other parameters, but at much less significance\footnote{As can be seen in the Appendix~\ref{App:S3}, the $S^3$ simulation predicts approximately the same number of SFGs and AGNs detected. Therefore, this effect is not present in the constraints obtained using $S^3$ as benchmark to predict the redshift distribution of galaxies and the galaxy and magnification bias.}. 

In Figure \ref{fig:contours} we show marginalized 68\% and 95\% confidence level contours on the $w_0$-$w_a$ plane from EMU forecasts (assuming the realization at design sensitivity and using two tracers), Planck+BAO and the combination of them (left panels), and contours on the $\mu_0$-$\gamma_0$ plane from EMU and  EMU combined with Planck+BAO+RSD in the corresponding right panels. Upper panels assume total knowledge of the galaxy and magnification bias, while bottom panels assume total ignorance and include a marginalization over $\Delta b_{\rm i}^X$ and $\Delta s_{\rm all}^X$. 
As can be seen, the degeneracy between $w_0$ and $w_a$ is almost the same measured by EMU and Planck+BAO, hence the combination of both measurements do not break the existing degeneracy. We only show one-dimensional marginalized 68\% constraints for $\mu_0$ and $\gamma_0$ independently from Planck+BAO+RSD as there is no published value for the correlation between $\mu_0$ and $\gamma_0$ nor a MCMC to compute it. 

The fact that an early EMU data release covers only 2000 deg$^2$ limits considerably its performance in constraining cosmological parameters with galaxy clustering. Therefore, combining Planck+BAO with EMU-early does not improve the constraints. However, the impact of having a factor two worse sensitivity than in the design sensitivity (i.e., the pessimistic sensitivity of 20 $\mu$Jy rms/beam) does not degrade critically the constraints. Using one single tracer and considering five redshift bins, the constraints are around $25-30\%$ weaker than at design sensitivity for all the parameters. In the case of $f_{\rm NL}$ using only one redshift bin, the constraints are even better, due to having a larger abundance ratio of AGNs with respect to SFGs, so the average bias is larger. 
 This is no longer true using five redshift bins, since the effect of the larger shot noise with the pessimistic sensitivity is more important in this case. Finally, the improvement that SKA-2 will achieve (a factor 10 in sensitivity) will be crucial. SKA-2 forecast constraints on $f_{\rm NL}$ assuming five redshift bins and a single tracer are a factor $\sim 3$ stronger than those for EMU at design sensitivity, improving upon current bounds from Planck. In addition, in the cases of evolving dark energy and modified gravity models, the constraints on $w_0$, $w_a$, $\mu_0$ and $\gamma_0$ will be a factor $\sim 2$ stronger than EMU at design sensitivity if ISW and galaxy clustering are combined. 

\begin{figure}
\includegraphics[width=\linewidth]{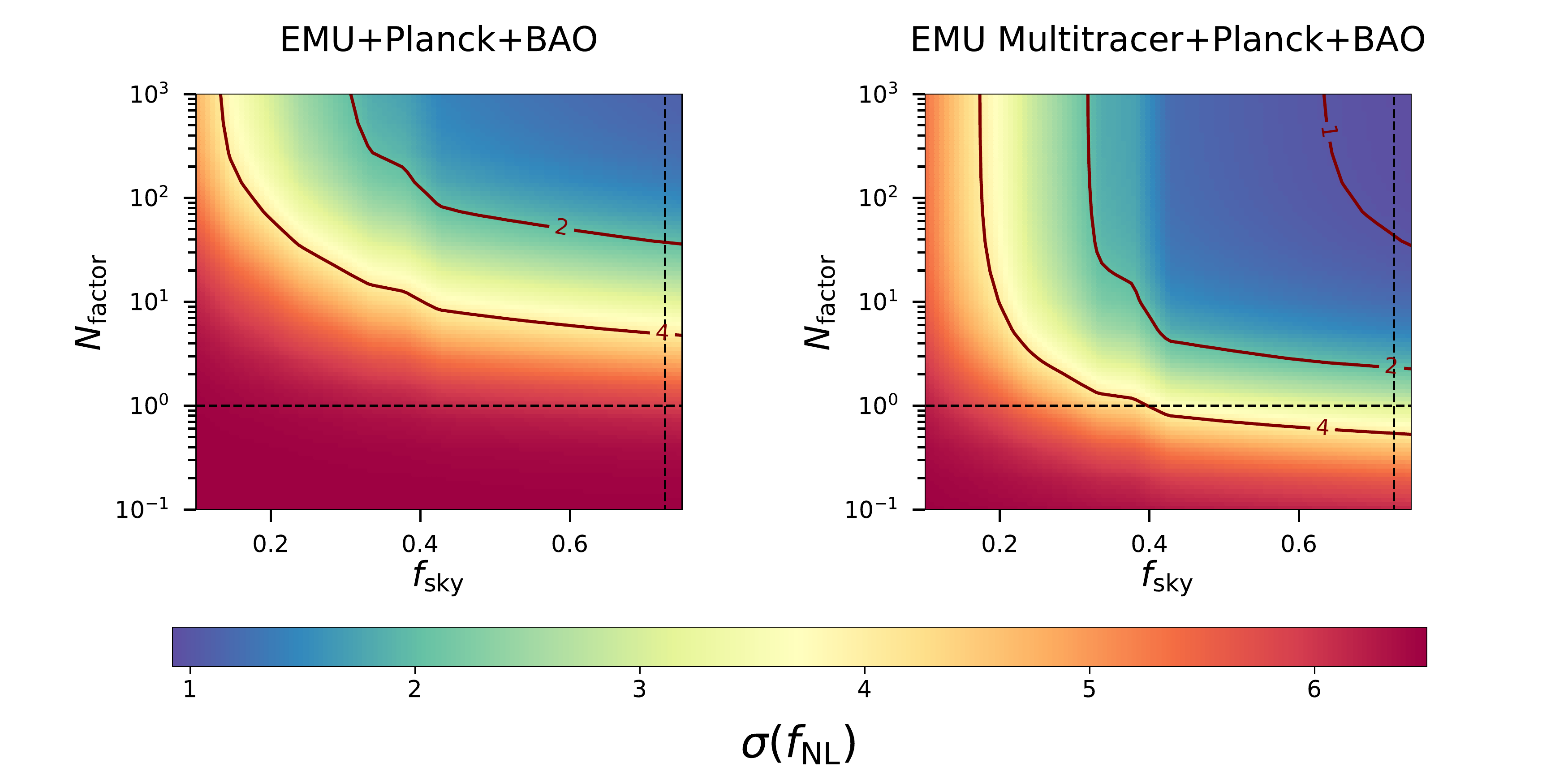}
\caption{68$\%$ confidence level forecast constraints for $f_{\rm NL}$ (assuming a fiducial $f_{\rm NL}=0$) from measurements of the angular galaxy power spectra and the ISW effect, combined with the prior from Planck and BAO. We use five redshift bins when considering all galaxies as one single tracer (left), but only one redshift bin when distinguishing between SFGs and AGNs (right) since we find better constraints than binning the catalogues in redshift (see main text and Table~\ref{tab:constr_multi}). }
\label{fig:fNL_var}
\end{figure}

However, EMU will not be able to measure $f_{\rm NL}$ below unity. 
In order to find the specifications needed by EMU to achieve the goal of having an uncertainty on local PNG measurements better than $\sigma(f_{\rm NL})\sim 1$, we forecast the constraints on $f_{\rm NL}$ from EMU-{\it like} surveys as a function of the fraction of sky surveyed and the number of sources detected.  
We model the variation on the number of sources with a factor $N_{\rm factor}$ multiplying the number density $dN/dzd\Omega$ appearing in Figure \ref{fig:specs}. We reckon that having a better sensitivity will allow to detect more new galaxies at larger redshift than at lower redshift, which would modify the shape of the redshift distribution. However, we expect this change in the shape of the redshift distribution to be small enough within the range of parameters considered here so as not to affect our results significantly. 
  Therefore, changing the total number of detected sources maintaining the shape of the redshift distribution is a fair approximation for this study. 
The variation to $f_{\rm sky}$ and $N_{\rm factor}$ enter  the covariance matrix in the Fisher matrix computation (Equations \eqref{eq:fisher_gg_tr} and \eqref{eq:fisher_isw}).  This kind of study is important in order to guide the planning of the survey strategy, since in principle it may be more convenient to use the same total observing time on a smaller area to reduce the noise and detect more sources, or vice versa.  For a similar investigation for spectroscopic surveys, see~\cite{Raccanelli:fNL}.

We show the dependence on $f_{\rm sky}$ and $N_{\rm factor}$ (starting from EMU at design sensitivity) for the predicted constraints on $f_{\rm NL}$ in Figure \ref{fig:fNL_var}, both considering all galaxies as the same kind of tracer (left panel) and using two tracers (right panel). In the multi-tracer case, we use a single redshift bin, since the constraints are better in this case. We use EMU forecasts assuming complete knowledge of the galaxy and magnification bias and the priors from current observations in both panels.  
We find that using all the galaxies as a single tracer, it will not be possible to measure local PNG with precision $\sigma(f_{\rm NL})<1$ even if EMU scans 75\% of the sky and observes 1000 times more galaxies. However, it will be possible using two different tracers if EMU would be able to detect  around 40 times more galaxies. In this the case, the total observing time of EMU should increase by a large factor in order to achieve this precision goal (note that reducing the rms/beam flux from 10 to 1 $\mu$Jy corresponds to detecting around 13 times more galaxies). On the other hand, using the same total observing time, the sensitivity can be increased if $f_{\rm sky}$ is smaller. However, we find that the $N_{\rm factor}$ needed to reach $\sigma(f_{\rm NL})\sim 1$ increases dramatically whenever $f_{\rm sky}<0.65$, given that, as stated above, PNG effects shows on ultra-large scales. Nonetheless, it is important to keep in mind that this analysis is done keeping the same abundance ratio between AGNs and SFGs. If, by any cause, a significant increase of sensitivity will amount to increasing the relative abundance of AGNs, or by any change, the measured galaxy bias is larger than expected, the total number of sources observed would not have to be so large. As can be seen in i.e. Tables~\ref{tab:constr_fLN} and~\ref{tab:constr_multi}, marginalizing over $\Delta b_{\rm i}^X$ and $\Delta s_{\rm i}^X$ does not have a large effect on the constraints on $f_{\rm NL}$, so the results are qualitatively similar in that case.

\section{Conclusions}\label{sec:discussion}
The Evolutionary Map of the Universe (EMU), an all-sky radio-continuum survey operating on the Australian SKA Pathfinder (ASKAP), will provide deep and wide observations with enough detected sources to study galaxy clustering at the ultra-large scales for the first time. In this work we 
 use the Fisher matrix formalism to estimate the precision in measuring parameters of models beyond $\Lambda$CDM, using measurements of the angular galaxy power spectrum and the integrated Sachs-Wolfe effect. The cases under study include models with evolving dark energy equation of state, low redshift deviations from General Relativity, spatial curvature and primordial non-Gaussianity in the local limit.

 We estimate the population of objects detected by EMU using the T-RECS catalogues~\cite{Bonaldi_TRECS} and assume that all observational systematics are under control and correctly accounted for. In this way, the main uncertainties left are the galaxy properties, such as the galaxy and magnification biases. We consider different levels of knowledge of these quantities throughout this work, having in mind that partial information will be available by the time  EMU data is available. Regarding the observed sources, we first consider all the galaxies as the same kind of tracer of the density field, and then distinguish between Star-Forming galaxies and AGNs as different tracers and perform a multi-tracer analysis. 

As a radio-continuum survey, EMU's redshift determination will be very poor. However, given  the large numbers and the improvement of the redshift inference algorithms and external data sets, it will be possible to assign redshifts so that a tomographic analysis can be performed. 
While not being competitive using all galaxies as the same tracer and a single redshift bin, we find that using external observations to infer the redshifts and split the catalogue into five redshift bins (which is a conservative assumption) boosts EMU's constraining power. Moreover, if the observations allow a reliable distinction between SFGs and AGNs, a multi-tracer analysis will return the best from EMU: when combined with current observations from CMB observations and BAO analyses, EMU's observations at design sensitivity will improve the current bounds on evolving dark energy models and modified gravity by a factor of two.

As EMU will survey a large fraction of the sky, it will observe the largest scales to date. Since local PNG would have imprints on the largest scales, EMU is a perfect experiment to increase the precision of the measurements of $f_{\rm NL}$. However, it will need a multi-tracer analysis in order to overcome the cosmic variance. This way, EMU alone will set a bound twice smaller than the current bound, and slightly smaller if combined with current observations. In order to achieve the possibly game-changing threshold of an uncertainty on $f_{\rm NL}$ below 1, EMU will have to observe around 40 times more galaxies, which would need an out-of-range amount of time of observations (i.e. SKA-2 would detect a factor 13 more galaxies than EMU). However, if the fraction of the sky observed lies below $\sim 0.65$ (which will allow to use more observation time per pointing to increase sensitivity), the amount of galaxies needed increases dramatically. 
Therefore, other strategies (combination with other cosmological probes, improvements in the redshift estimation with external data such it is possible to split the catalogue in more redshift bins, increasing the number of tracers, etc.) will be needed to achieve this goal. Note that using $S^3$ instead of T-RECS the constraints are slightly better (see Appendix~\ref{App:S3}). Nonetheless, even in this case, achieving $\sigma(f_{\rm NL})<1$ with only EMU seems difficult and it will be needed to wait for more sensitive experiments such as the different surveys of SKA-2. This result is consistent with the findings reported for the SKA Phase 1~\cite{redbook}. However, the possibility to find small deviations from Gaussian initial conditions with SKA will only be possible if it follows an all-sky survey strategy. In any case, EMU's bound on $f_{\rm NL}$ will be the best in the near future and EMU's catalogues will be the main data set to combine with new measurements in the quest to measure PNG.

Although EMU is operating in ASKAP, a pathfinder for SKA, it will have scientific relevance on its own, since it will be the deepest all-sky survey (and with best angular resolution among similar surveys) by the time it will end. This will be an important step forward to constrain physics which manifests itselfs on the largest scales, such as primordial non-Gaussianity, but also relativistic corrections in the galaxy clustering statistics \cite{Borzyszkowski_liger}. This makes EMU a critical stepping stone for our understanding of the early Universe, gravity and dark energy, and also for the preparation of future galaxy surveys.

\acknowledgments
We thank Nicola Bellomo for discussions during the development of this work. 
Funding
for this work was partially provided by the Spanish MINECO under
projects AYA2014-58747-P AEI/FEDER UE and MDM-2014-0369 of ICCUB
(Unidad de Excelencia Maria de Maeztu).  JLB is supported by the
Spanish MINECO under grant BES-2015-071307, co-funded by the ESF. 
AR has received funding from the People Programme (Marie Curie Actions) of the European Union H2020 Programme under REA grant agreement number 706896 (COSMOFLAGS) and the John Templeton Foundation. This work was supported at Johns Hopkins by NSF Grant No. 0244990, NASA NNX17AK38G, and the Templeton foundation.

\bibliography{biblio}

\providecommand{\href}[2]{#2}\begingroup\raggedright\begin{thebibliography}{10}

\bibitem{Planck18_pars}
{Planck Collaboration}, N.~{Aghanim}, {\em et~al.}, ``{Planck 2018 results. VI.
  Cosmological parameters},'' {\em ArXiv e-prints} (July, 2018) ,
  \href{http://arxiv.org/abs/1807.06209}{{\ttfamily arXiv:1807.06209}}.

\bibitem{Alam_bossdr12}
{\bfseries SDSS-III BOSS} Collaboration, S.~{Alam} and et~al., ``{The
  clustering of galaxies in the completed SDSS-III Baryon Oscillation
  Spectroscopic Survey: cosmological analysis of the DR12 galaxy sample},''
  \href{http://dx.doi.org/10.1093/mnras/stx721}{{\em MNRAS} {\bfseries 470}
  (Sept., 2017) 2617--2652}, \href{http://arxiv.org/abs/1607.03155}{{\ttfamily
  arXiv:1607.03155}}.

\bibitem{RiessH0_2016}
A.~G. {Riess}, L.~M. {Macri}, S.~L. {Hoffmann}, D.~{Scolnic}, S.~{Casertano},
  A.~V. {Filippenko}, B.~E. {Tucker}, M.~J. {Reid}, D.~O. {Jones}, J.~M.
  {Silverman}, R.~{Chornock}, P.~{Challis}, W.~{Yuan}, P.~J. {Brown}, and R.~J.
  {Foley}, ``{A 2.4\% Determination of the Local Value of the Hubble
  Constant},'' \href{http://dx.doi.org/10.3847/0004-637X/826/1/56}{{\em ApJ}
  {\bfseries 826} (July, 2016) 56},
  \href{http://arxiv.org/abs/1604.01424}{{\ttfamily arXiv:1604.01424}}.

\bibitem{Riess18_GaiaDR2}
A.~G. {Riess} {\em et~al.}, ``{Milky Way Cepheid Standards for Measuring Cosmic
  Distances and Application to Gaia DR2: Implications for the Hubble
  Constant},'' \href{http://dx.doi.org/10.3847/1538-4357/aac82e}{{\em
  Astrophys. J.} {\bfseries 861} (July, 2018) 126},
  \href{http://arxiv.org/abs/1804.10655}{{\ttfamily arXiv:1804.10655}}.

\bibitem{BernalH0}
J.~L. {Bernal}, L.~{Verde}, and A.~G. {Riess}, ``{The trouble with H$_{0}$},''
  \href{http://dx.doi.org/10.1088/1475-7516/2016/10/019}{{\em JCAP} {\bfseries
  10} (Oct., 2016) 019}, \href{http://arxiv.org/abs/1607.05617}{{\ttfamily
  arXiv:1607.05617}}.

\bibitem{Bernal_baccus}
J.~L. {Bernal} and J.~A. {Peacock}, ``{Conservative cosmology: combining data
  with allowance for unknown systematics},''
  \href{http://dx.doi.org/10.1088/1475-7516/2018/07/002}{{\em JCAP} {\bfseries
  7} (July, 2018) 002}, \href{http://arxiv.org/abs/1803.04470}{{\ttfamily
  arXiv:1803.04470}}.

\bibitem{Poulin_H0}
V.~{Poulin}, K.~K. {Boddy}, S.~{Bird}, and M.~{Kamionkowski}, ``{Implications
  of an extended dark energy cosmology with massive neutrinos for cosmological
  tensions},'' \href{http://dx.doi.org/10.1103/PhysRevD.97.123504}{{\em Phys.
  Rev D} {\bfseries 97} (June, 2018) 123504},
  \href{http://arxiv.org/abs/1803.02474}{{\ttfamily arXiv:1803.02474
  [astro-ph.CO]}}.

\bibitem{DiValentino_interDE}
E.~{Di Valentino}, A.~{Melchiorri}, and O.~{Mena}, ``{Can interacting dark
  energy solve the H$_{0}$ tension?},''
  \href{http://dx.doi.org/10.1103/PhysRevD.96.043503}{{\em Phys. Rev D}
  {\bfseries 96} no.~4, (Aug., 2017) 043503},
  \href{http://arxiv.org/abs/1704.08342}{{\ttfamily arXiv:1704.08342}}.

\bibitem{DiValentino:2016hlg}
E.~{Di Valentino}, A.~{Melchiorri}, and J.~{Silk}, ``{Reconciling Planck with
  the local value of H$_{0}$ in extended parameter space},''
  \href{http://dx.doi.org/10.1016/j.physletb.2016.08.043}{{\em Physics Letters
  B} {\bfseries 761} (Oct., 2016) 242--246},
  \href{http://arxiv.org/abs/1606.00634}{{\ttfamily arXiv:1606.00634}}.

\bibitem{Deramo_H0}
F.~{D'Eramo}, R.~Z. {Ferreira}, A.~{Notari}, and J.~L. {Bernal}, ``{Hot axions
  and the H$_{0}$ tension},''
  \href{http://dx.doi.org/10.1088/1475-7516/2018/11/014}{{\em Journal of
  Cosmology and Astro-Particle Physics} {\bfseries 2018} (Nov., 2018) 014},
  \href{http://arxiv.org/abs/1808.07430}{{\ttfamily arXiv:1808.07430
  [hep-ph]}}.

\bibitem{NVSS}
J.~J. {Condon}, W.~D. {Cotton}, E.~W. {Greisen}, Q.~F. {Yin}, R.~A. {Perley},
  G.~B. {Taylor}, and J.~J. {Broderick}, ``{The NRAO VLA Sky Survey},''
  \href{http://dx.doi.org/10.1086/300337}{{\em Astrophys. J} {\bfseries 115}
  (May, 1998) 1693--1716}.

\bibitem{Boughn_radio}
S.~P. {Boughn} and R.~G. {Crittenden}, ``{Cross Correlation of the Cosmic
  Microwave Background with Radio Sources: Constraints on an Accelerating
  Universe},'' \href{http://dx.doi.org/10.1103/PhysRevLett.88.021302}{{\em
  Physical Review Letters} {\bfseries 88} no.~2, (Jan., 2002) 021302},
  \href{http://arxiv.org/abs/astro-ph/0111281}{{\ttfamily astro-ph/0111281}}.

\bibitem{Overzier_nvssclust}
R.~A. {Overzier}, H.~J.~A. {R{\"o}ttgering}, R.~B. {Rengelink}, and R.~J.
  {Wilman}, ``{The spatial clustering of radio sources in NVSS and FIRST;
  implications for galaxy clustering evolution},''
  \href{http://dx.doi.org/10.1051/0004-6361:20030527}{{\em Astronom. \&
  Astrophys.} {\bfseries 405} (July, 2003) 53--72},
  \href{http://arxiv.org/abs/astro-ph/0304160}{{\ttfamily astro-ph/0304160}}.

\bibitem{Boughn_isw}
S.~{Boughn} and R.~{Crittenden}, ``{A correlation between the cosmic microwave
  background and large-scale structure in the Universe},''
  \href{http://dx.doi.org/10.1038/nature02139}{{\em Nature} {\bfseries 427}
  (Jan., 2004) 45--47}, \href{http://arxiv.org/abs/astro-ph/0305001}{{\ttfamily
  astro-ph/0305001}}.

\bibitem{Nolta_isw}
M.~R. {Nolta}, E.~L. {Wright}, L.~{Page}, C.~L. {Bennett}, M.~{Halpern},
  G.~{Hinshaw}, N.~{Jarosik}, A.~{Kogut}, M.~{Limon}, S.~S. {Meyer}, D.~N.
  {Spergel}, G.~S. {Tucker}, and E.~{Wollack}, ``{First Year Wilkinson
  Microwave Anisotropy Probe Observations: Dark Energy Induced Correlation with
  Radio Sources},'' \href{http://dx.doi.org/10.1086/386536}{{\em Astrophys. J.}
  {\bfseries 608} (June, 2004) 10--15},
  \href{http://arxiv.org/abs/astro-ph/0305097}{{\ttfamily astro-ph/0305097}}.

\bibitem{Smith_radiolensing}
K.~M. {Smith}, O.~{Zahn}, and O.~{Dor{\'e}}, ``{Detection of gravitational
  lensing in the cosmic microwave background},''
  \href{http://dx.doi.org/10.1103/PhysRevD.76.043510}{{\em Phys. Rev. D}
  {\bfseries 76} no.~4, (Aug., 2007) 043510},
  \href{http://arxiv.org/abs/0705.3980}{{\ttfamily arXiv:0705.3980}}.

\bibitem{Raccanelli_radioisw}
A.~{Raccanelli}, A.~{Bonaldi}, M.~{Negrello}, S.~{Matarrese}, G.~{Tormen}, and
  G.~{de Zotti}, ``{A reassessment of the evidence of the Integrated
  Sachs-Wolfe effect through the WMAP-NVSS correlation},''
  \href{http://dx.doi.org/10.1111/j.1365-2966.2008.13189.x}{{\em MNRAS}
  {\bfseries 386} (June, 2008) 2161--2166},
  \href{http://arxiv.org/abs/0802.0084}{{\ttfamily arXiv:0802.0084}}.

\bibitem{Ho_nvssisw}
S.~{Ho}, C.~{Hirata}, N.~{Padmanabhan}, U.~{Seljak}, and N.~{Bahcall},
  ``{Correlation of CMB with large-scale structure. I. Integrated Sachs-Wolfe
  tomography and cosmological implications},''
  \href{http://dx.doi.org/10.1103/PhysRevD.78.043519}{{\em Phys. Rev. D}
  {\bfseries 78} no.~4, (Aug., 2008) 043519},
  \href{http://arxiv.org/abs/0801.0642}{{\ttfamily arXiv:0801.0642}}.

\bibitem{Afshordi_fNL}
N.~{Afshordi} and A.~J. {Tolley}, ``{Primordial non-Gaussianity, statistics of
  collapsed objects, and the integrated Sachs-Wolfe effect},''
  \href{http://dx.doi.org/10.1103/PhysRevD.78.123507}{{\em Phys. Rev. D}
  {\bfseries 78} no.~12, (Dec., 2008) 123507},
  \href{http://arxiv.org/abs/0806.1046}{{\ttfamily arXiv:0806.1046}}.

\bibitem{Xia_gammaray}
J.-Q. {Xia}, A.~{Cuoco}, E.~{Branchini}, M.~{Fornasa}, and M.~{Viel}, ``{A
  cross-correlation study of the Fermi-LAT {$\gamma$}-ray diffuse extragalactic
  signal},'' \href{http://dx.doi.org/10.1111/j.1365-2966.2011.19200.x}{{\em
  MNRAS} {\bfseries 416} (Sept., 2011) 2247--2264},
  \href{http://arxiv.org/abs/1103.4861}{{\ttfamily arXiv:1103.4861
  [astro-ph.CO]}}.

\bibitem{Rubart_dipole}
M.~{Rubart} and D.~J. {Schwarz}, ``{Cosmic radio dipole from NVSS and WENSS},''
  \href{http://dx.doi.org/10.1051/0004-6361/201321215}{{\em Astronom. \&
  Astrophys.} {\bfseries 555} (July, 2013) A117},
  \href{http://arxiv.org/abs/1301.5559}{{\ttfamily arXiv:1301.5559
  [astro-ph.CO]}}.

\bibitem{Giannantonio_isw13}
T.~{Giannantonio}, A.~J. {Ross}, W.~J. {Percival}, R.~{Crittenden},
  D.~{Bacher}, M.~{Kilbinger}, R.~{Nichol}, and J.~{Weller}, ``{Improved
  primordial non-Gaussianity constraints from measurements of galaxy clustering
  and the integrated Sachs-Wolfe effect},''
  \href{http://dx.doi.org/10.1103/PhysRevD.89.023511}{{\em Phys. Rev. D}
  {\bfseries 89} no.~2, (Jan., 2014) 023511},
  \href{http://arxiv.org/abs/1303.1349}{{\ttfamily arXiv:1303.1349
  [astro-ph.CO]}}.

\bibitem{Nusser_radioclust}
A.~{Nusser} and P.~{Tiwari}, ``{The Clustering of Radio Galaxies: Biasing and
  Evolution versus Stellar Mass},''
  \href{http://dx.doi.org/10.1088/0004-637X/812/1/85}{{\em Astrophys. J.}
  {\bfseries 812} (Oct., 2015) 85},
  \href{http://arxiv.org/abs/1505.06817}{{\ttfamily arXiv:1505.06817}}.

\bibitem{Planck15_isw}
{Planck Collaboration}, P.~A.~R. {Ade}, N.~{Aghanim}, M.~{Arnaud},
  M.~{Ashdown}, J.~{Aumont}, C.~{Baccigalupi}, A.~J. {Banday}, R.~B.
  {Barreiro}, N.~{Bartolo}, and et~al., ``{Planck 2015 results. XXI. The
  integrated Sachs-Wolfe effect},''
  \href{http://dx.doi.org/10.1051/0004-6361/201525831}{{\em Astronom. \&
  Astrophys.} {\bfseries 594} (Sept., 2016) A21},
  \href{http://arxiv.org/abs/1502.01595}{{\ttfamily arXiv:1502.01595}}.

\bibitem{Raccanelli_gw}
A.~{Raccanelli}, ``{Gravitational wave astronomy with radio galaxy surveys},''
  \href{http://dx.doi.org/10.1093/mnras/stx835}{{\em MNRAS} {\bfseries 469}
  (July, 2017) 656--670}, \href{http://arxiv.org/abs/1609.09377}{{\ttfamily
  arXiv:1609.09377}}.

\bibitem{Raccanelli_radio}
A.~{Raccanelli}, G.-B. {Zhao}, D.~J. {Bacon}, M.~J. {Jarvis}, W.~J. {Percival},
  R.~P. {Norris}, H.~{R{\"o}ttgering}, F.~B. {Abdalla}, C.~M. {Cress}, J.-C.
  {Kubwimana}, S.~{Lindsay}, R.~C. {Nichol}, M.~G. {Santos}, and D.~J.
  {Schwarz}, ``{Cosmological measurements with forthcoming radio continuum
  surveys},'' \href{http://dx.doi.org/10.1111/j.1365-2966.2012.20634.x}{{\em
  MNRAS} {\bfseries 424} (Aug., 2012) 801--819},
  \href{http://arxiv.org/abs/1108.0930}{{\ttfamily arXiv:1108.0930}}.

\bibitem{Camera_radio}
S.~{Camera}, M.~G. {Santos}, D.~J. {Bacon}, M.~J. {Jarvis}, K.~{McAlpine},
  R.~P. {Norris}, A.~{Raccanelli}, and H.~{R{\"o}ttgering}, ``{Impact of
  redshift information on cosmological applications with next-generation radio
  surveys},'' \href{http://dx.doi.org/10.1111/j.1365-2966.2012.22073.x}{{\em
  MNRAS} {\bfseries 427} (Dec., 2012) 2079--2088},
  \href{http://arxiv.org/abs/1205.1048}{{\ttfamily arXiv:1205.1048}}.

\bibitem{Raccanelli_iswfNL}
A.~{Raccanelli}, O.~{Dor{\'e}}, D.~J. {Bacon}, R.~{Maartens}, M.~G. {Santos},
  S.~{Camera}, T.~M. {Davis}, M.~J. {Drinkwater}, M.~{Jarvis}, R.~{Norris}, and
  D.~{Parkinson}, ``{Probing primordial non-Gaussianity via iSW measurements
  with SKA continuum surveys},''
  \href{http://dx.doi.org/10.1088/1475-7516/2015/01/042}{{\em JCAP} {\bfseries
  1} (Jan., 2015) 042}, \href{http://arxiv.org/abs/1406.0010}{{\ttfamily
  arXiv:1406.0010}}.

\bibitem{Bertacca_udm}
D.~{Bertacca}, A.~{Raccanelli}, O.~F. {Piattella}, D.~{Pietrobon},
  N.~{Bartolo}, S.~{Matarrese}, and T.~{Giannantonio}, ``{CMB-galaxy
  correlation in Unified Dark Matter scalar field cosmologies},''
  \href{http://dx.doi.org/10.1088/1475-7516/2011/03/039}{{\em JCAP} {\bfseries
  3} (Mar., 2011) 039}, \href{http://arxiv.org/abs/1102.0284}{{\ttfamily
  arXiv:1102.0284}}.

\bibitem{Jarvis_SKA}
M.~{Jarvis}, D.~{Bacon}, C.~{Blake}, M.~{Brown}, S.~{Lindsay}, A.~{Raccanelli},
  M.~{Santos}, and D.~J. {Schwarz}, ``{Cosmology with SKA Radio Continuum
  Surveys},'' {\em Advancing Astrophysics with the Square Kilometre Array
  (AASKA14)} (Apr., 2015) 18, \href{http://arxiv.org/abs/1501.03825}{{\ttfamily
  arXiv:1501.03825}}.

\bibitem{Camera_fNL}
S.~{Camera}, M.~G. {Santos}, and R.~{Maartens}, ``{Probing primordial
  non-Gaussianity with SKA galaxy redshift surveys: a fully relativistic
  analysis},'' \href{http://dx.doi.org/10.1093/mnras/stv040}{{\em MNRAS}
  {\bfseries 448} (Apr., 2015) 1035--1043},
  \href{http://arxiv.org/abs/1409.8286}{{\ttfamily arXiv:1409.8286}}.

\bibitem{Ferramacho_radioPNG}
L.~D. {Ferramacho}, M.~G. {Santos}, M.~J. {Jarvis}, and S.~{Camera}, ``{Radio
  galaxy populations and the multitracer technique: pushing the limits on
  primordial non-Gaussianity},''
  \href{http://dx.doi.org/10.1093/mnras/stu1015}{{\em MNRAS} {\bfseries 442}
  (Aug., 2014) 2511--2518}.

\bibitem{Raccanelli2017_PNG}
A.~{Raccanelli}, M.~{Shiraishi}, N.~{Bartolo}, D.~{Bertacca}, M.~{Liguori},
  S.~{Matarrese}, R.~P. {Norris}, and D.~{Parkinson}, ``{Future constraints on
  angle-dependent non-Gaussianity from large radio surveys},''
  \href{http://dx.doi.org/10.1016/j.dark.2016.10.006}{{\em Physics of the Dark
  Universe} {\bfseries 15} (Mar., 2017) 35--46}.

\bibitem{Karagiannis_PNG}
D.~{Karagiannis}, A.~{Lazanu}, M.~{Liguori}, A.~{Raccanelli}, N.~{Bartolo}, and
  L.~{Verde}, ``{Constraining primordial non-Gaussianity with bispectrum and
  power spectrum from upcoming optical and radio surveys},''
  \href{http://dx.doi.org/10.1093/mnras/sty1029}{{\em MNRAS} {\bfseries 478}
  (July, 2018) 1341--1376}, \href{http://arxiv.org/abs/1801.09280}{{\ttfamily
  arXiv:1801.09280}}.

\bibitem{Scelfo_progenitors}
G.~{Scelfo}, N.~{Bellomo}, A.~{Raccanelli}, S.~{Matarrese}, and L.~{Verde},
  ``{GW{$\times$}LSS: chasing the progenitors of merging binary black holes},''
  \href{http://dx.doi.org/10.1088/1475-7516/2018/09/039}{{\em JCAP} {\bfseries
  9} (Sept., 2018) 039}, \href{http://arxiv.org/abs/1809.03528}{{\ttfamily
  arXiv:1809.03528}}.

\bibitem{Ballardini_primordial}
M.~{Ballardini}, F.~{Finelli}, R.~{Maartens}, and L.~{Moscardini}, ``{Probing
  primordial features with next-generation photometric and radio surveys},''
  \href{http://dx.doi.org/10.1088/1475-7516/2018/04/044}{{\em JCAP} {\bfseries
  4} (Apr., 2018) 044}, \href{http://arxiv.org/abs/1712.07425}{{\ttfamily
  arXiv:1712.07425}}.

\bibitem{Alonso_ST}
D.~{Alonso}, E.~{Bellini}, P.~G. {Ferreira}, and M.~{Zumalac{\'a}rregui},
  ``{Observational future of cosmological scalar-tensor theories},''
  \href{http://dx.doi.org/10.1103/PhysRevD.95.063502}{{\em Phys. Rev. D}
  {\bfseries 95} no.~6, (Mar., 2017) 063502},
  \href{http://arxiv.org/abs/1610.09290}{{\ttfamily arXiv:1610.09290}}.

\bibitem{EMU}
R.~P. {Norris} {\em et~al.}, ``{EMU: Evolutionary Map of the Universe},''
  \href{http://dx.doi.org/10.1071/AS11021}{{\em PASA} {\bfseries 28} (Aug.,
  2011) 215--248}, \href{http://arxiv.org/abs/1106.3219}{{\ttfamily
  arXiv:1106.3219}}.

\bibitem{Johnston_askap07}
S.~{Johnston} {\em et~al.}, ``{Science with the Australian Square Kilometre
  Array Pathfinder},'' \href{http://dx.doi.org/10.1071/AS07033}{{\em PASA}
  {\bfseries 24} (Dec., 2007) 174--188}.

\bibitem{Johnston_askap08}
S.~{Johnston} {\em et~al.}, ``Science with ASKAP. The Australian
  square-kilometre-array pathfinder,''
  \href{http://dx.doi.org/10.1007/s10686-008-9124-7}{{\em Experimental
  Astronomy} {\bfseries 22} (Dec., 2008) 151--273}.

\bibitem{Norris_ASKAP}
R.~P. {Norris} {\em et~al.}, ``{Radio Continuum Surveys with Square Kilometre
  Array Pathfinders},'' \href{http://dx.doi.org/10.1017/pas.2012.020}{{\em
  PASA} {\bfseries 30} (Mar., 2013) e020},
  \href{http://arxiv.org/abs/1210.7521}{{\ttfamily arXiv:1210.7521}}.

\bibitem{Bartolo_png}
N.~{Bartolo}, E.~{Komatsu}, S.~{Matarrese}, and A.~{Riotto}, ``{Non-Gaussianity
  from inflation: theory and observations},''
  \href{http://dx.doi.org/10.1016/j.physrep.2004.08.022}{{\em Physical Reports}
  {\bfseries 402} (Nov., 2004) 103--266},
  \href{http://arxiv.org/abs/astro-ph/0406398}{{\ttfamily astro-ph/0406398}}.

\bibitem{Komatsu_png}
E.~{Komatsu}, ``{Hunting for primordial non-Gaussianity in the cosmic microwave
  background},'' \href{http://dx.doi.org/10.1088/0264-9381/27/12/124010}{{\em
  Classical and Quantum Gravity} {\bfseries 27} no.~12, (June, 2010) 124010},
  \href{http://arxiv.org/abs/1003.6097}{{\ttfamily arXiv:1003.6097}}.

\bibitem{Wands_png}
D.~{Wands}, ``{Local non-Gaussianity from inflation},''
  \href{http://dx.doi.org/10.1088/0264-9381/27/12/124002}{{\em Classical and
  Quantum Gravity} {\bfseries 27} no.~12, (June, 2010) 124002},
  \href{http://arxiv.org/abs/1004.0818}{{\ttfamily arXiv:1004.0818}}.

\bibitem{Alvarez_fNL}
M.~{Alvarez}, T.~{Baldauf}, J.~R. {Bond}, N.~{Dalal}, R.~{de Putter},
  O.~{Dor{\'e}}, D.~{Green}, C.~{Hirata}, Z.~{Huang}, D.~{Huterer}, D.~{Jeong},
  M.~C. {Johnson}, E.~{Krause}, M.~{Loverde}, J.~{Meyers}, P.~D. {Meerburg},
  L.~{Senatore}, S.~{Shandera}, E.~{Silverstein}, A.~{Slosar}, K.~{Smith},
  M.~{Zaldarriaga}, V.~{Assassi}, J.~{Braden}, A.~{Hajian}, T.~{Kobayashi},
  G.~{Stein}, and A.~{van Engelen}, ``{Testing Inflation with Large Scale
  Structure: Connecting Hopes with Reality},'' {\em ArXiv e-prints} (Dec.,
  2014) , \href{http://arxiv.org/abs/1412.4671}{{\ttfamily arXiv:1412.4671}}.

\bibitem{Norris_MLredshift}
R.~P. {Norris}, M.~{Salvato}, and G.~{Longo}, ``{},'' {\em Submitted to PASP} .

\bibitem{Luken_MLredshift}
K.~J. {Luken}, R.~P. {Norris}, and L.~A.~F. {Park}, ``{Preliminary results of
  using k-Nearest Neighbours Regression to estimate the redshift of radio
  selected datasets},'' {\em arXiv e-prints} (Oct., 2018) arXiv:1810.10714,
  \href{http://arxiv.org/abs/1810.10714}{{\ttfamily arXiv:1810.10714
  [astro-ph.GA]}}.

\bibitem{Menard_cbr}
B.~{M{\'e}nard}, R.~{Scranton}, S.~{Schmidt}, C.~{Morrison}, D.~{Jeong},
  T.~{Budavari}, and M.~{Rahman}, ``{Clustering-based redshift estimation:
  method and application to data},'' {\em ArXiv e-prints} (Mar., 2013) ,
  \href{http://arxiv.org/abs/1303.4722}{{\ttfamily arXiv:1303.4722}}.

\bibitem{Rahman_cbr}
M.~{Rahman}, B.~{M{\'e}nard}, R.~{Scranton}, S.~J. {Schmidt}, and C.~B.
  {Morrison}, ``{Clustering-based redshift estimation: comparison to
  spectroscopic redshifts},''
  \href{http://dx.doi.org/10.1093/mnras/stu2636}{{\em MNRAS} {\bfseries 447}
  (Mar., 2015) 3500--3511}, \href{http://arxiv.org/abs/1407.7860}{{\ttfamily
  arXiv:1407.7860}}.

\bibitem{Kovetz_cbr}
E.~D. {Kovetz}, A.~{Raccanelli}, and M.~{Rahman}, ``{Cosmological constraints
  with clustering-based redshifts},''
  \href{http://dx.doi.org/10.1093/mnras/stx691}{{\em MNRAS} {\bfseries 468}
  (July, 2017) 3650--3656}, \href{http://arxiv.org/abs/1606.07434}{{\ttfamily
  arXiv:1606.07434}}.

\bibitem{Bonaldi_TRECS}
A.~{Bonaldi}, M.~{Bonato}, V.~{Galluzzi}, I.~{Harrison}, M.~{Massardi},
  S.~{Kay}, G.~{De Zotti}, and M.~L. {Brown}, ``{The Tiered Radio Extragalactic
  Continuum Simulation (T-RECS)},''
  \href{http://dx.doi.org/10.1093/mnras/sty2603}{{\em MNRAS} {\bfseries 482}
  (Jan., 2019) 2--19}, \href{http://arxiv.org/abs/1805.05222}{{\ttfamily
  arXiv:1805.05222 [astro-ph.GA]}}.

\bibitem{Wilman_S3}
R.~J. {Wilman}, L.~{Miller}, M.~J. {Jarvis}, T.~{Mauch}, F.~{Levrier}, F.~B.
  {Abdalla}, S.~{Rawlings}, H.~R. {Kl{\"o}ckner}, D.~{Obreschkow},
  D.~{Olteanu}, and S.~{Young}, ``{A semi-empirical simulation of the
  extragalactic radio continuum sky for next generation radio telescopes},''
  \href{http://dx.doi.org/10.1111/j.1365-2966.2008.13486.x}{{\em MNRAS}
  {\bfseries 388} (Aug., 2008) 1335--1348}.

\bibitem{Seljak_fNLmultitracer}
U.~{Seljak}, ``{Extracting Primordial Non-Gaussianity without Cosmic
  Variance},'' \href{http://dx.doi.org/10.1103/PhysRevLett.102.021302}{{\em
  Physical Review Letters} {\bfseries 102} no.~2, (Jan., 2009) 021302},
  \href{http://arxiv.org/abs/0807.1770}{{\ttfamily arXiv:0807.1770}}.

\bibitem{McDonald_multitracer}
P.~{McDonald} and U.~{Seljak}, ``{How to evade the sample variance limit on
  measurements of redshift-space distortions},''
  \href{http://dx.doi.org/10.1088/1475-7516/2009/10/007}{{\em JCAP} {\bfseries
  10} (Oct., 2009) 007}, \href{http://arxiv.org/abs/0810.0323}{{\ttfamily
  arXiv:0810.0323}}.

\bibitem{Raccanelli_radiosurveys}
A.~{Raccanelli}, G.-B. {Zhao}, D.~J. {Bacon}, M.~J. {Jarvis}, W.~J. {Percival},
  R.~P. {Norris}, H.~{R{\"o}ttgering}, F.~B. {Abdalla}, C.~M. {Cress}, J.-C.
  {Kubwimana}, S.~{Lindsay}, R.~C. {Nichol}, M.~G. {Santos}, and D.~J.
  {Schwarz}, ``{Cosmological measurements with forthcoming radio continuum
  surveys},'' \href{http://dx.doi.org/10.1111/j.1365-2966.2012.20634.x}{{\em
  MNRAS} {\bfseries 424} (Aug., 2012) 801--819}.

\bibitem{Matsubara_lensing2000}
T.~Matsubara, ``{The gravitational lensing in redshift-space correlation
  functions of galaxies and quasars},''
  \href{http://dx.doi.org/10.1086/312762}{{\em Astrophys. J.} {\bfseries 537}
  (2000) L77},
\href{http://arxiv.org/abs/astro-ph/0004392}{{\ttfamily arXiv:astro-ph/0004392
  [astro-ph]}}.
%%CITATION = ASTRO-PH/0004392;%%.

\bibitem{Bartelmann_mag01}
M.~{Bartelmann} and P.~{Schneider}, ``{Weak gravitational lensing},''
  \href{http://dx.doi.org/10.1016/S0370-1573(00)00082-X}{{\em Phyics Reporst}
  {\bfseries 340} (Jan., 2001) 291--472}.

\bibitem{Liu_mag14}
J.~Liu, Z.~Haiman, L.~Hui, J.~M. Kratochvil, and M.~May, ``The Impact of
  Magnification and Size Bias on Weak Lensing Power Spectrum and Peak
  Statistics,'' {\em Phys. Rev. D} {\bfseries 89} (2014) 023515,
  \href{http://arxiv.org/abs/1310.7517}{{\ttfamily 1310.7517}}.
  \url{https://arxiv.org/abs/1310.7517}.

\bibitem{Bellomo_Fisher}
N.~{Bellomo}, J.~L. {Bernal}, A.~{Raccanelli}, L.~{Verde}, and G.~{Scelfo},
  ``{},'' {\em In prep.} .

\bibitem{Lesgourgues:2011re}
J.~Lesgourgues, ``{The Cosmic Linear Anisotropy Solving System (CLASS) I:
  Overview},''
\href{http://arxiv.org/abs/1104.2932}{{\ttfamily arXiv:1104.2932
  [astro-ph.IM]}}.
%%CITATION = ARXIV:1104.2932;%%.

\bibitem{Yoo_GR09}
J.~{Yoo}, A.~L. {Fitzpatrick}, and M.~{Zaldarriaga}, ``{New perspective on
  galaxy clustering as a cosmological probe: General relativistic effects},''
  \href{http://dx.doi.org/10.1103/PhysRevD.80.083514}{{\em Phys. Rev. D}
  {\bfseries 80} no.~8, (Oct., 2009) 083514},
  \href{http://arxiv.org/abs/0907.0707}{{\ttfamily arXiv:0907.0707
  [astro-ph.CO]}}.

\bibitem{Yoo_GR10}
J.~{Yoo}, ``{General relativistic description of the observed galaxy power
  spectrum: Do we understand what we measure?},''
  \href{http://dx.doi.org/10.1103/PhysRevD.82.083508}{{\em Phys. Rev. D}
  {\bfseries 82} no.~8, (Oct., 2010) 083508},
  \href{http://arxiv.org/abs/1009.3021}{{\ttfamily arXiv:1009.3021}}.

\bibitem{Bonvin_obsLSS}
C.~{Bonvin} and R.~{Durrer}, ``{What galaxy surveys really measure},''
  \href{http://dx.doi.org/10.1103/PhysRevD.84.063505}{{\em Phys. Rev. D}
  {\bfseries 84} (Sept., 2011) 063505}.

\bibitem{Challinor_obsLSS}
A.~{Challinor} and A.~{Lewis}, ``{Linear power spectrum of observed source
  number counts},'' \href{http://dx.doi.org/10.1103/PhysRevD.84.043516}{{\em
  Phys. Rev. D} {\bfseries 84} (Aug., 2011) 043516}.

\bibitem{Jeong_GR12}
D.~{Jeong}, F.~{Schmidt}, and C.~M. {Hirata}, ``{Large-scale clustering of
  galaxies in general relativity},''
  \href{http://dx.doi.org/10.1103/PhysRevD.85.023504}{{\em Phys. Rev. D}
  {\bfseries 85} no.~2, (Jan., 2012) 023504},
  \href{http://arxiv.org/abs/1107.5427}{{\ttfamily arXiv:1107.5427}}.

\bibitem{Bertacca_xi3dGR}
D.~{Bertacca}, R.~{Maartens}, A.~{Raccanelli}, and C.~{Clarkson}, ``{Beyond the
  plane-parallel and Newtonian approach: wide-angle redshift distortions and
  convergence in general relativity},''
  \href{http://dx.doi.org/10.1088/1475-7516/2012/10/025}{{\em JCAP} {\bfseries
  10} (Oct., 2012) 025}, \href{http://arxiv.org/abs/1205.5221}{{\ttfamily
  arXiv:1205.5221}}.

\bibitem{Raccanelli_GRCl}
A.~{Raccanelli}, F.~{Montanari}, D.~{Bertacca}, O.~{Dor{\'e}}, and R.~{Durrer},
  ``{Cosmological measurements with general relativistic galaxy
  correlations},'' \href{http://dx.doi.org/10.1088/1475-7516/2016/05/009}{{\em
  JCAP} {\bfseries 2016} (May, 2016) 009}.

\bibitem{Raccanelli_GRint}
A.~{Raccanelli}, D.~{Bertacca}, R.~{Maartens}, C.~{Clarkson}, and
  O.~{Dor{\'e}}, ``{Lensing and time-delay contributions to galaxy
  correlations},'' \href{http://dx.doi.org/10.1007/s10714-016-2076-8}{{\em
  General Relativity and Gravitation} {\bfseries 48} (July, 2016) 84}.

\bibitem{Raccanelli_doppler}
A.~{Raccanelli}, D.~{Bertacca}, D.~{Jeong}, M.~C. {Neyrinck}, and A.~S.
  {Szalay}, ``{Doppler term in the galaxy two-point correlation function:
  Wide-angle, velocity, Doppler lensing and cosmic acceleration effects},''
  \href{http://dx.doi.org/10.1016/j.dark.2017.12.003}{{\em Physics of the Dark
  Universe} {\bfseries 19} (Mar., 2018) 109--123},
  \href{http://arxiv.org/abs/1602.03186}{{\ttfamily arXiv:1602.03186}}.

\bibitem{DiDio_Classgal}
E.~{Di Dio}, F.~{Montanari}, J.~{Lesgourgues}, and R.~{Durrer}, ``{The CLASSgal
  code for relativistic cosmological large scale structure},''
  \href{http://dx.doi.org/10.1088/1475-7516/2013/11/044}{{\em JCAP} {\bfseries
  11} (Nov., 2013) 044}, \href{http://arxiv.org/abs/1307.1459}{{\ttfamily
  arXiv:1307.1459}}.

\bibitem{Raccanelli_radial}
A.~{Raccanelli}, D.~{Bertacca}, R.~{Maartens}, C.~{Clarkson}, and
  O.~{Dor{\'e}}, ``{Lensing and time-delay contributions to galaxy
  correlations},'' \href{http://dx.doi.org/10.1007/s10714-016-2076-8}{{\em
  General Relativity and Gravitation} {\bfseries 48} (July, 2016) 84},
  \href{http://arxiv.org/abs/1311.6813}{{\ttfamily arXiv:1311.6813}}.

\bibitem{Giannantonio_isw08}
T.~{Giannantonio}, R.~{Scranton}, R.~G. {Crittenden}, R.~C. {Nichol}, S.~P.
  {Boughn}, A.~D. {Myers}, and G.~T. {Richards}, ``{Combined analysis of the
  integrated Sachs-Wolfe effect and cosmological implications},''
  \href{http://dx.doi.org/10.1103/PhysRevD.77.123520}{{\em Phys. Rev. D}
  {\bfseries 77} no.~12, (June, 2008) 123520},
  \href{http://arxiv.org/abs/0801.4380}{{\ttfamily arXiv:0801.4380}}.

\bibitem{Afshordi_isw}
N.~{Afshordi}, Y.-S. {Loh}, and M.~A. {Strauss}, ``{Cross-correlation of the
  cosmic microwave background with the 2MASS galaxy survey: Signatures of dark
  energy, hot gas, and point sources},''
  \href{http://dx.doi.org/10.1103/PhysRevD.69.083524}{{\em Phys. Rev. D}
  {\bfseries 69} no.~8, (Apr., 2004) 083524},
  \href{http://arxiv.org/abs/astro-ph/0308260}{{\ttfamily astro-ph/0308260}}.

\bibitem{Fosalba_isw}
P.~{Fosalba} and E.~{Gazta{\~n}aga}, ``{Measurement of the gravitational
  potential evolution from the cross-correlation between WMAP and the APM
  Galaxy Survey},''
  \href{http://dx.doi.org/10.1111/j.1365-2966.2004.07837.x}{{\em MNRAS}
  {\bfseries 350} (May, 2004) L37--L41},
  \href{http://arxiv.org/abs/astro-ph/0305468}{{\ttfamily astro-ph/0305468}}.

\bibitem{Scranton_isw}
R.~{Scranton}, A.~J. {Connolly}, R.~C. {Nichol}, A.~{Stebbins}, I.~{Szapudi},
  D.~J. {Eisenstein}, N.~{Afshordi}, T.~{Budavari}, I.~{Csabai}, J.~A.
  {Frieman}, J.~E. {Gunn}, D.~{Johnston}, Y.~{Loh}, R.~H. {Lupton}, C.~J.
  {Miller}, E.~S. {Sheldon}, R.~S. {Sheth}, A.~S. {Szalay}, M.~{Tegmark}, and
  Y.~{Xu}, ``{Physical Evidence for Dark Energy},'' {\em ArXiv Astrophysics
  e-prints} (July, 2003) ,
  \href{http://arxiv.org/abs/astro-ph/0307335}{{\ttfamily astro-ph/0307335}}.

\bibitem{Xia_PNG}
J.-Q. {Xia}, A.~{Bonaldi}, C.~{Baccigalupi}, G.~{De Zotti}, S.~{Matarrese},
  L.~{Verde}, and M.~{Viel}, ``{Constraining primordial non-Gaussianity with
  high-redshift probes},''
  \href{http://dx.doi.org/10.1088/1475-7516/2010/08/013}{{\em JCAP} {\bfseries
  2010} (Aug., 2010) 013}.

\bibitem{Matarrese_png00}
S.~{Matarrese}, L.~{Verde}, and R.~{Jimenez}, ``{The Abundance of High-Redshift
  Objects as a Probe of Non-Gaussian Initial Conditions},''
  \href{http://dx.doi.org/10.1086/309412}{{\em Astrophys. J.} {\bfseries 541}
  (Sept., 2000) 10--24},
  \href{http://arxiv.org/abs/astro-ph/0001366}{{\ttfamily astro-ph/0001366}}.

\bibitem{Dalal_png07}
N.~{Dalal}, O.~{Dor{\'e}}, D.~{Huterer}, and A.~{Shirokov}, ``{Imprints of
  primordial non-Gaussianities on large-scale structure: Scale-dependent bias
  and abundance of virialized objects},''
  \href{http://dx.doi.org/10.1103/PhysRevD.77.123514}{{\em Phys. Rev. D}
  {\bfseries 77} no.~12, (June, 2008) 123514},
  \href{http://arxiv.org/abs/0710.4560}{{\ttfamily arXiv:0710.4560}}.

\bibitem{Matarrese_png08}
S.~{Matarrese} and L.~{Verde}, ``{The Effect of Primordial Non-Gaussianity on
  Halo Bias},'' \href{http://dx.doi.org/10.1086/587840}{{\em Astrophys. J.
  Letters} {\bfseries 677} (Apr., 2008) L77},
  \href{http://arxiv.org/abs/0801.4826}{{\ttfamily arXiv:0801.4826}}.

\bibitem{Desjacques_png}
V.~{Desjacques} and U.~{Seljak}, ``{Primordial non-Gaussianity from the
  large-scale structure},''
  \href{http://dx.doi.org/10.1088/0264-9381/27/12/124011}{{\em Classical and
  Quantum Gravity} {\bfseries 27} no.~12, (June, 2010) 124011},
  \href{http://arxiv.org/abs/1003.5020}{{\ttfamily arXiv:1003.5020}}.

\bibitem{CPL_w0wa}
M.~{Chevallier} and D.~{Polarski}, ``{Accelerating Universes with Scaling Dark
  Matter},'' \href{http://dx.doi.org/10.1142/S0218271801000822}{{\em
  International Journal of Modern Physics D} {\bfseries 10} (Jan., 2001)
  213--223}.

\bibitem{Linder_w0wa}
E.~V. {Linder}, ``{Exploring the Expansion History of the Universe},''
  \href{http://dx.doi.org/10.1103/PhysRevLett.90.091301}{{\em Physical Review
  Letters} {\bfseries 90} no.~9, (Mar., 2003) 091301},
  \href{http://arxiv.org/abs/astro-ph/0208512}{{\ttfamily astro-ph/0208512}}.

\bibitem{Ezquiaga_GWreview}
J.~M. Ezquiaga and M.~Zumalacárregui, ``{Dark Energy in light of
  Multi-Messenger Gravitational-Wave astronomy},''
  \href{http://dx.doi.org/10.3389/fspas.2018.00044}{{\em Front. Astron. Space
  Sci.} {\bfseries 5} (2018) 44},
\href{http://arxiv.org/abs/1807.09241}{{\ttfamily arXiv:1807.09241
  [astro-ph.CO]}}.
%%CITATION = ARXIV:1807.09241;%%.

\bibitem{Bernal_ParamSplit}
J.~L. {Bernal}, L.~{Verde}, and A.~J. {Cuesta}, ``{Parameter splitting in dark
  energy: is dark energy the same in the background and in the cosmic
  structures?},'' \href{http://dx.doi.org/10.1088/1475-7516/2016/02/059}{{\em
  JCAP} {\bfseries 2} (Feb., 2016) 059},
  \href{http://arxiv.org/abs/1511.03049}{{\ttfamily arXiv:1511.03049}}.

\bibitem{Amendola_MG}
L.~{Amendola}, M.~{Kunz}, and D.~{Sapone}, ``{Measuring the dark side (with
  weak lensing)},'' \href{http://dx.doi.org/10.1088/1475-7516/2008/04/013}{{\em
  JCAP} {\bfseries 4} (Apr., 2008) 013},
  \href{http://arxiv.org/abs/0704.2421}{{\ttfamily arXiv:0704.2421}}.

\bibitem{Zhao_MG}
G.-B. {Zhao}, T.~{Giannantonio}, L.~{Pogosian}, A.~{Silvestri}, D.~J. {Bacon},
  K.~{Koyama}, R.~C. {Nichol}, and Y.-S. {Song}, ``{Probing modifications of
  general relativity using current cosmological observations},''
  \href{http://dx.doi.org/10.1103/PhysRevD.81.103510}{{\em Phys. Rev. D}
  {\bfseries 81} no.~10, (May, 2010) 103510},
  \href{http://arxiv.org/abs/1003.0001}{{\ttfamily arXiv:1003.0001
  [astro-ph.CO]}}.

\bibitem{Baker_mgclass}
T.~{Baker} and P.~{Bull}, ``{Observational Signatures of Modified Gravity on
  Ultra-large Scales},''
  \href{http://dx.doi.org/10.1088/0004-637X/811/2/116}{{\em Astrophys. J.}
  {\bfseries 811} (Oct., 2015) 116}.

\bibitem{Fisher:1935}
R.~A. Fisher, ``{The Fiducial Argument in Statistical Inference},''
\href{http://dx.doi.org/10.1111/j.1469-1809.1935.tb02120.x}{{\em Annals Eugen.}
  {\bfseries 6} (1935) 391--398}.
%%CITATION = ANEUA,6,391;%%.

\bibitem{Tegmark_fisher97}
M.~{Tegmark}, A.~N. {Taylor}, and A.~F. {Heavens}, ``{Karhunen-Lo{\`e}ve
  Eigenvalue Problems in Cosmology: How Should We Tackle Large Data Sets?},''
  \href{http://dx.doi.org/10.1086/303939}{{\em Astrophys. J.} {\bfseries 480}
  (May, 1997) 22--35}, \href{http://arxiv.org/abs/astro-ph/9603021}{{\ttfamily
  astro-ph/9603021}}.

\bibitem{Verde_wmap03}
L.~{Verde}, H.~V. {Peiris}, D.~N. {Spergel}, M.~R. {Nolta}, C.~L. {Bennett},
  M.~{Halpern}, G.~{Hinshaw}, N.~{Jarosik}, A.~{Kogut}, M.~{Limon}, S.~S.
  {Meyer}, L.~{Page}, G.~S. {Tucker}, E.~{Wollack}, and E.~L. {Wright},
  ``{First-Year Wilkinson Microwave Anisotropy Probe (WMAP) Observations:
  Parameter Estimation Methodology},''
  \href{http://dx.doi.org/10.1086/377335}{{\em The Astrophysical Journal
  Supplement Series} {\bfseries 148} (Sept., 2003) 195--211}.

\bibitem{Sellentin_dali}
E.~{Sellentin}, M.~{Quartin}, and L.~{Amendola}, ``{Breaking the spell of
  Gaussianity: forecasting with higher order Fisher matrices},''
  \href{http://dx.doi.org/10.1093/mnras/stu689}{{\em MNRAS} {\bfseries 441}
  (June, 2014) 1831--1840}, \href{http://arxiv.org/abs/1401.6892}{{\ttfamily
  arXiv:1401.6892}}.

\bibitem{Beutler11}
F.~{Beutler}, C.~{Blake}, M.~{Colless}, D.~H. {Jones}, L.~{Staveley-Smith},
  L.~{Campbell}, Q.~{Parker}, W.~{Saunders}, and F.~{Watson}, ``{The 6dF Galaxy
  Survey: baryon acoustic oscillations and the local Hubble constant},''
  \href{http://dx.doi.org/10.1111/j.1365-2966.2011.19250.x}{{\em Mon. Not. Roy.
  Astron. Soc.} {\bfseries 416} (Oct., 2011) 3017--3032},
  \href{http://arxiv.org/abs/1106.3366}{{\ttfamily arXiv:1106.3366}}.

\bibitem{Ross15}
A.~J. {Ross}, L.~{Samushia}, C.~{Howlett}, W.~J. {Percival}, A.~{Burden}, and
  M.~{Manera}, ``{The clustering of the SDSS DR7 main Galaxy sample - I. A 4
  per cent distance measure at z = 0.15},''
  \href{http://dx.doi.org/10.1093/mnras/stv154}{{\em Mon. Not. Roy. Astron.
  Soc.} {\bfseries 449} (May, 2015) 835--847},
  \href{http://arxiv.org/abs/1409.3242}{{\ttfamily arXiv:1409.3242}}.

\bibitem{Planck15_PNG}
{Planck Collaboration}, P.~A.~R. {Ade}, and others., ``{Planck 2015 results.
  XVII. Constraints on primordial non-Gaussianity},''
  \href{http://dx.doi.org/10.1051/0004-6361/201525836}{{\em A\&A} {\bfseries
  594} (Sept., 2016) A17}, \href{http://arxiv.org/abs/1502.01592}{{\ttfamily
  arXiv:1502.01592}}.

\bibitem{Raccanelli:fNL}
A.~{Raccanelli}, O.~{Dor{\'e}}, and N.~{Dalal}, ``{Optimization of
  spectroscopic surveys for testing non-Gaussianity},''
  \href{http://dx.doi.org/10.1088/1475-7516/2015/08/034}{{\em JCAP} {\bfseries
  8} (Aug., 2015) 034}, \href{http://arxiv.org/abs/1409.1927}{{\ttfamily
  arXiv:1409.1927}}.

\bibitem{redbook}
{Square Kilometre Array Cosmology Science Working Group}, D.~J. {Bacon}, {\em
  et~al.}, ``{Cosmology with Phase 1 of the Square Kilometre Array; Red Book
  2018: Technical specifications and performance forecasts},'' {\em arXiv
  e-prints} (Nov., 2018) arXiv:1811.02743,
  \href{http://arxiv.org/abs/1811.02743}{{\ttfamily arXiv:1811.02743
  [astro-ph.CO]}}.

\bibitem{Borzyszkowski_liger}
M.~{Borzyszkowski}, D.~{Bertacca}, and C.~{Porciani}, ``{liger: mock
  relativistic light cones from Newtonian simulations},''
  \href{http://dx.doi.org/10.1093/mnras/stx1423}{{\em MNRAS} {\bfseries 471}
  (Nov., 2017) 3899--3914}, \href{http://arxiv.org/abs/1703.03407}{{\ttfamily
  arXiv:1703.03407}}.

\end{thebibliography}\endgroup
\bibliographystyle{utcaps} 

\appendix

\section{Cosmological forecasts results}
\label{App:tables}
In this appendix we report the results of the cosmological forecast for radio-continuum surveys discussed in Section \ref{sec:results} in detail (Tables~\ref{tab:constr_fLN},~\ref{tab:constr_w0wa},~\ref{tab:constr_omegak},~\ref{tab:constr_MG} and~\ref{tab:constr_multi}. We consider the four models discussed in Section \ref{sec:models} and four different surveys: EMU at design sensitivity (10 $\mu$Jy rms flux/beam), a pessimistic realization (with twice rms flux/beam), EMU early results (100 $\mu$Jy rms flux/beam) and SKA-2 (1 $\mu$Jy rms flux/beam). For each survey, we consider a single redshift bin and five different redshift bins, as shown in Section~\ref{sec:EMU}. Moreover, we consider different assumptions about our prior knowledge for the galaxy and magnification biases. Either we understand the biases completely, we marginalize over a single parameter which shifts $b(z)$, marginalize over parameters which shifts $b(z)$ for each bin or marginalize over parameters that shift both $b(z)$ and $s(z)$ in each bin.  

% \usepackage{multirow}
\begin{table}[]
\centering
\begin{tabular}{|c|c|c|c|c|c|}
\hline
\multirow{2}{*}{Survey} & \multirow{2}{*}{\begin{tabular}[c]{@{}c@{}}\# redshift \\ bins\end{tabular}} & \multirow{2}{*}{\begin{tabular}[c]{@{}c@{}}Bias \\ uncertainty\end{tabular}} & \multicolumn{3}{c|}{\begin{tabular}[c]{@{}c@{}}Data combination and constraints on \\ $\Lambda$CDM+$f_{\rm NL}$\end{tabular}} \\ \cline{4-6} 
 &  &  & Galaxy Clustering (GC) & GC+ISW & GC+ISW+Planck+BAO \\ \hline
\multicolumn{3}{|c|}{} & $f_{\rm NL}$ & $f_{\rm NL}$ & $f_{\rm NL}$ \\ \hline
\multirow{7}{*}{\begin{tabular}[c]{@{}c@{}}EMU\\ \\ Design \\ Sensitivity\\ (10 $\mu$Jy\\  rms/beam)\end{tabular}} & \multirow{3}{*}{1 bin} & Known & 240 & 140 & 6.4 \\
 &  & $\Delta b_{\rm all}$ & 320 & 170 & 6.4 \\
 &  & $\Delta b_{\rm all}$ \& $\Delta s_{\rm all}$ & 690 & 240 & 6.4 \\ \cline{2-6} 
 & \multirow{4}{*}{5 bins} & Known & 17 & 16 & 5.9 \\
 &  & $\Delta b_{\rm all}$ & 17 & 16 & 5.9 \\
 &  & $\Delta b_i$ & 17 & 16 & 5.9 \\
 &  & $\Delta b_i$ \& $\Delta s_i$ & 19 & 18 & 5.9 \\ \hline
\multirow{7}{*}{\begin{tabular}[c]{@{}c@{}}EMU\\ \\ Pessimistic \\ Sensitivity\\ (20 $\mu$Jy\\  rms/beam)\end{tabular}} & \multirow{3}{*}{1 bin} & Known & 160 & 130 & 6.4 \\
 &  & $\Delta b_{\rm all}$ & 360 & 140 & 6.4 \\
 &  & $\Delta b_{\rm all}$ \& $\Delta s_{\rm all}$ & 360 & 160 & 6.5 \\ \cline{2-6} 
 & \multirow{4}{*}{5 bins} & Known & 23 & 21 & 6.1 \\
 &  & $\Delta b_{\rm all}$ & 23 & 21 & 6.1 \\
 &  & $\Delta b_i$ & 24 & 22 & 6.1 \\
 &  & $\Delta b_i$ \& $\Delta s_i$ & 27 & 24 & 6.1 \\ \hline
\multirow{7}{*}{\begin{tabular}[c]{@{}c@{}}EMU\\ \\ Early\\ Results\\ (100 $\mu$Jy \\ rms/beam,\\ 2000 deg$^2$)\end{tabular}} & \multirow{3}{*}{1 bin} & Known & 6200 & 3000 & 6.5 \\
 &  & $\Delta b_{\rm all}$ & 50000 & 4000 & 6.5 \\
 &  & $\Delta b_{\rm all}$ \& $\Delta s_{\rm all}$ & 74000 & 4400 & 6.5 \\ \cline{2-6} 
 & \multirow{4}{*}{5 bins} & Known & 360 & 360 & 6.5 \\
 &  & $\Delta b_{\rm all}$ & 360 & 360 & 6.5 \\
 &  & $\Delta b_i$ & 400 & 390 & 6.5 \\
 &  & $\Delta b_i$ \& $\Delta s_i$ & 490 & 470 & 6.5 \\ \hline
\multirow{7}{*}{\begin{tabular}[c]{@{}c@{}}SKA-2\\ \\ (1 $\mu$Jy \\ rms/beam)\end{tabular}} & \multirow{3}{*}{1 bin} & Known & 130 & 70 & 6.4 \\
 &  & $\Delta b_{\rm all}$ & 130 & 70 & 6.4 \\
 &  & $\Delta b_{\rm all}$ \& $\Delta s_{\rm all}$ & 270 & 73 & 6.4 \\ \cline{2-6} 
 & \multirow{4}{*}{5 bins} & Known & 5.5 & 5.5 & 4.0 \\
 &  & $\Delta b_{\rm all}$ & 5.6 & 5.5 & 4.0 \\
 &  & $\Delta b_i$ & 5.7 & 5.6 & 4.1 \\
 &  & $\Delta b_i$ \& $\Delta s_i$ & 5.9 & 5.8 & 4.1 \\ \hline
\end{tabular}
\caption{Marginalized  68\%  confidence  level  predicted  constraints  on  $f_{\rm NL}$ assuming $\Lambda$CDM+$f_{\rm NL}$ from radio-continuum measurements with EMU.  We  show  results  of  galaxy  clustering  by  itself  (GC),  galaxy  clustering combined  with  ISW  (GC+ISW),  and  when  Planck+BAO  priors  are  also  added  (GC+ISW+Planck+BAO),both not  binning  in  redshift and with five redshift bins. We consider four different surveys: EMU at design sensitivity, a pessimistic realization of EMU with twice the rms flux/beam, EMU early and SKA-2. We assume different cases for the knowledge of the bias, either known or marginalizing over the galaxy bias and the magnification bias. $\Delta b_{\rm all}$ refers to marginalizing over a single parameter $\Delta b_all$ which shifts the whole $b(z)$.}
\label{tab:constr_fLN}
\end{table}

% \usepackage{multirow}
\begin{table}[]
\centering
\begin{tabular}{|c|c|c|c|c|c|c|c|c|}
\hline
\multirow{2}{*}{Survey} & \multirow{2}{*}{\begin{tabular}[c]{@{}c@{}}\# redshift\\  bins\end{tabular}} & \multirow{2}{*}{\begin{tabular}[c]{@{}c@{}}Bias \\ uncertainty\end{tabular}} & \multicolumn{6}{c|}{\begin{tabular}[c]{@{}c@{}}Data combination and constraints on \\ ($w_0w_a$)CDM\end{tabular}} \\ \cline{4-9} 
 &  &  & \multicolumn{2}{c|}{GC} & \multicolumn{2}{c|}{GC+ISW} & \multicolumn{2}{c|}{GC+ISW+Planck+BAO} \\ \hline
\multicolumn{3}{|c|}{} & $w_0$ & $w_a$ & $w_0$ & $w_a$ & $w_0$ & $w_a$ \\ \hline
\multirow{7}{*}{\begin{tabular}[c]{@{}c@{}}EMU\\ \\ Design \\ Sensitivity\\ (10 $\mu$Jy \\ rms/beam)\end{tabular}} & \multirow{3}{*}{1 bin} & Known & 12 & 34 & 2.3 & 7.2 & 0.20 & 0.48 \\
 &  & $\Delta b_{\rm all}$ & 19 & 70 & 5.0 & 9.6 & 0.26 & 0.72 \\
 &  & $\Delta b_{\rm all}$ \& $\Delta s_{\rm all}$ & 61 & 210 & 5.3 & 12 & 0.26 & 0.74 \\ \cline{2-9} 
 & \multirow{4}{*}{5 bins} & Known & 0.66 & 1.9 & 0.59 & 1.7 & 0.20 & 0.50 \\
 &  & $\Delta b_{\rm all}$ & 0.78 & 2.9 & 0.60 & 1.9 & 0.23 & 0.66 \\
 &  & $\Delta b_i$ & 1.8 & 6.5 & 1.0 & 2.9 & 0.25 & 0.69 \\
 &  & $\Delta b_i$ \& $\Delta s_i$ & 2.2 & 8.1 & 1.4 & 3.5 & 0.26 & 0.71 \\ \hline
\multirow{7}{*}{\begin{tabular}[c]{@{}c@{}}EMU\\ \\ Pessimistic \\ Sensitivity\\ (20 $\mu$Jy rms/beam)\end{tabular}} & \multirow{3}{*}{1 bin} & Known & 6.9 & 23 & 2.5 & 8.3 & 0.21 & 0.51 \\
 &  & $\Delta b_{\rm all}$ & 29 & 120 & 3.5 & 8.7 & 0.25 & 0.71 \\
 &  & $\Delta b_{\rm all}$ \& $\Delta s_{\rm all}$ & 65 & 230 & 3.5 & 9.4 & 0.26 & 0.72 \\ \cline{2-9} 
 & \multirow{4}{*}{5 bins} & Known & 0.95 & 2.9 & 0.80 & 2.4 & 0.22 & 0.58 \\
 &  & $\Delta b_{\rm all}$ & 1.0 & 3.9 & 0.81 & 2.5 & 0.24 & 0.68 \\
 &  & $\Delta b_i$ & 2.4 & 9.9 & 1.2 & 3.5 & 0.25 & 0.71 \\
 &  & $\Delta b_i$ \& $\Delta s_i$ & 3.2 & 13 & 2.0 & 4.8 & 0.26 & 0.72 \\ \hline
\multirow{7}{*}{\begin{tabular}[c]{@{}c@{}}EMU\\ \\ Early\\ Resuls\\ (100 $\mu$Jy \\ rms/beam,\\ 2000 deg$^2$)\end{tabular}} & \multirow{3}{*}{1 bin} & Known & 2100 & 6600 & 51 & 130 & 0.26 & 0.73 \\
 &  & $\Delta b_{\rm all}$ & 2800 & 9900 & 270 & 550 & 0.27 & 0.74 \\
 &  & $\Delta b_{\rm all}$ \& $\Delta s_{\rm all}$ & 5400 & 33000 & 280 & 580 & 0.27 & 0.74 \\ \cline{2-9} 
 & \multirow{4}{*}{5 bins} & Known & 8.3 & 26 & 6.3 & 17 & 0.27 & 0.74 \\
 &  & $\Delta b_{\rm all}$ & 9.1 & 35 & 8.8 & 18 & 0.27 & 0.74 \\
 &  & $\Delta b_i$ & 28 & 100 & 16 & 36 & 0.27 & 0.74 \\
 &  & $\Delta b_i$ \& $\Delta s_i$ & 58 & 170 & 20 & 44 & 0.27 & 0.74 \\ \hline
\multirow{7}{*}{\begin{tabular}[c]{@{}c@{}}SKA-2\\ \\ \\ (1 $\mu$Jy \\ rms/beam)\end{tabular}} & \multirow{3}{*}{1 bin} & Known & 18 & 63 & 2.8 & 7.3 & 0.18 & 0.43 \\
 &  & $\Delta b_{\rm all}$ & 21 & 94 & 3.2 & 7.7 & 0.27 & 0.74 \\
 &  & $\Delta b_{\rm all}$ \& $\Delta s_{\rm all}$ & 22 & 120 & 7.2 & 16 & 0.27 & 0.74 \\ \cline{2-9} 
 & \multirow{4}{*}{5 bins} & Known & 0.33 & 0.87 & 0.31 & 0.80 & 0.15 & 0.36 \\
 &  & $\Delta b_{\rm all}$ & 0.44 & 1.3 & 0.37 & 1.1 & 0.21 & 0.60 \\
 &  & $\Delta b_i$ & 0.79 & 2.2 & 0.61 & 1.6 & 0.24 & 0.64 \\
 &  & $\Delta b_i$ \& $\Delta s_i$ & 0.94 & 2.9 & 0.68 & 1.7 & 0.24 & 0.66 \\ \hline
\end{tabular}
\caption{Same as Table~\ref{tab:constr_fLN} but for $w_0$ and $w_a$ assuming a $(w_0w_a)$CDM model.}
\label{tab:constr_w0wa}
\end{table}

% \usepackage{multirow}
\begin{table}[]
\centering
\begin{tabular}{|c|c|c|c|c|c|}
\hline
\multirow{2}{*}{Survey} & \multirow{2}{*}{\begin{tabular}[c]{@{}c@{}}\# redshift \\ bins\end{tabular}} & \multirow{2}{*}{\begin{tabular}[c]{@{}c@{}}Bias \\ uncertainty\end{tabular}} & \multicolumn{3}{c|}{\begin{tabular}[c]{@{}c@{}}Data combination and constraints on \\ $\Lambda$CDM+$\Omega_k$\end{tabular}} \\ \cline{4-6} 
 &  &  & GC & GC+ISW & GC+ISW+Planck+BAO \\ \hline
\multicolumn{3}{|c|}{} & $100\times \Omega_k$ & $100\times \Omega_k$ & $100\times \Omega_k$ \\ \hline
\multirow{7}{*}{\begin{tabular}[c]{@{}c@{}}EMU\\ \\ Design \\ Sensitivity\\ (10 $\mu$Jy \\ rms/beam)\end{tabular}} & \multirow{3}{*}{1 bin} & Known & 8.7 & 8.3 & 0.16 \\
 &  & $\Delta b_{\rm all}$ & 9.1 & 8.4 & 0.19 \\
 &  & $\Delta b_{\rm all}$ \& $\Delta s_{\rm all}$ & 9.1 & 8.4 & 0.19 \\ \cline{2-6} 
 & \multirow{4}{*}{5 bins} & Known & 6.3 & 4.8 & 0.17 \\
 &  & $\Delta b_{\rm all}$ & 9.6 & 7.8 & 0.19 \\
 &  & $\Delta b_i$ & 13 & 8.0 & 0.19 \\
 &  & $\Delta b_i$ \& $\Delta s_i$ & 15 & 8.8 & 0.19 \\ \hline
\multirow{7}{*}{\begin{tabular}[c]{@{}c@{}}EMU\\ \\ Pessimistic \\ Sensitivity\\ (20 $\mu$Jy \\ rms/beam)\end{tabular}} & \multirow{3}{*}{1 bin} & Known & 12 & 11 & 0.17 \\
 &  & $\Delta b_{\rm all}$ & 13 & 11 & 0.19 \\
 &  & $\Delta b_{\rm all}$ \& $\Delta s_{\rm all}$ & 13 & 11 & 0.19 \\ \cline{2-6} 
 & \multirow{4}{*}{5 bins} & Known & 9.7 & 6.7 & 0.18 \\
 &  & $\Delta b_{\rm all}$ & 18 & 12 & 0.19 \\
 &  & $\Delta b_i$ & 23 & 12 & 0.19 \\
 &  & $\Delta b_i$ \& $\Delta s_i$ & 28 & 13 & 0.19 \\ \hline
\multirow{7}{*}{\begin{tabular}[c]{@{}c@{}}EMU\\ \\ Early\\ Results\\ (100 $\mu$Jy\\ rms/beam,\\ 2000 deg$^2$)\end{tabular}} & \multirow{3}{*}{1 bin} & Known & 130 & 88 & 0.19 \\
 &  & $\Delta b_{\rm all}$ & 130 & 110 & 0.19 \\
 &  & $\Delta b_{\rm all}$ \& $\Delta s_{\rm all}$ & 130 & 110 & 0.19 \\ \cline{2-6} 
 & \multirow{4}{*}{5 bins} & Known & 87 & 45 & 0.19 \\
 &  & $\Delta b_{\rm all}$ & 150 & 77 & 0.19 \\
 &  & $\Delta b_i$ & 160 & 82 & 0.19 \\
 &  & $\Delta b_i$ \& $\Delta s_i$ & 160 & 82 & 0.19 \\ \hline
\multirow{7}{*}{\begin{tabular}[c]{@{}c@{}}SKA-2\\ \\ (1 $\mu$Jy \\ rms/beam)\end{tabular}} & \multirow{3}{*}{1 bin} & Known & 5.4 & 4.4 & 0.16 \\
 &  & $\Delta b_{\rm all}$ & 6.0 & 4.9 & 0.19 \\
 &  & $\Delta b_{\rm all}$ \& $\Delta s_{\rm all}$ & 6.0 & 5.3 & 0.19 \\ \cline{2-6} 
 & \multirow{4}{*}{5 bins} & Known & 2.3 & 2.0 & 0.15 \\
 &  & $\Delta b_{\rm all}$ & 2.6 & 2.4 & 0.18 \\
 &  & $\Delta b_i$ & 3.8 & 2.9 & 0.18 \\
 &  & $\Delta b_i$ \& $\Delta s_i$ & 4.3 & 3.1 & 0.18 \\ \hline
\end{tabular}
\caption{Same as Table~\ref{tab:constr_fLN} but for $\Omega_k$ assuming a $\Lambda$CDM+$\Omega_k$ model.}
\label{tab:constr_omegak}
\end{table}

% \usepackage{multirow}
\begin{table}[]
\centering
\begin{tabular}{|c|c|c|c|c|c|c|c|c|}
\hline
\multirow{2}{*}{Survey} & \multirow{2}{*}{\begin{tabular}[c]{@{}c@{}}\# redshift \\ bins\end{tabular}} & \multirow{2}{*}{\begin{tabular}[c]{@{}c@{}}Bias \\ uncertainty\end{tabular}} & \multicolumn{6}{c|}{\begin{tabular}[c]{@{}c@{}}Data combination and constraints on \\ $\Lambda$CDM+$\mu_0$+$\gamma_0$\end{tabular}} \\ \cline{4-9} 
 &  &  & \multicolumn{2}{c|}{GC} & \multicolumn{2}{c|}{GC+ISW} & \multicolumn{2}{c|}{GC+ISW+Planck+BAO+RSD} \\ \hline
\multicolumn{3}{|c|}{} & $\mu_0$ & $\gamma_0$ & $\mu_0$ & $\gamma_0$ & $\mu_0$ & $\gamma_0$ \\ \hline
\multirow{7}{*}{\begin{tabular}[c]{@{}c@{}}EMU\\ \\ Design \\ Sensitivity\\ (10 $\mu$Jy \\ rms/beam)\end{tabular}} & \multirow{3}{*}{1 bin} & Known &  23 & 46 & 0.84 & 1.9 & 0.14 & 0.35 \\
 &  & $\Delta b_{\rm all}$ & 25 & 51 & 1.0 & 1.9 & 0.20 & 0.44 \\
 &  & $\Delta b_{\rm all}$ \& $\Delta s_{\rm all}$ & 32 & 63 & 1.0 & 1.9 & 0.20 & 0.45 \\ \cline{2-9} 
 & \multirow{4}{*}{5 bins} & Known & 0.64 & 1.9 & 0.34 & 0.81 & 0.12 & 0.30 \\
 &  & $\Delta b_{\rm all}$ & 1.1 & 2.8 & 0.43 & 0.95 & 0.18 & 0.40 \\
 &  & $\Delta b_i$ & 3.9 & 8.8 & 0.46 & 1.0 & 0.19 & 0.42 \\
 &  & $\Delta b_i$ \& $\Delta s_i$ & 5.2 & 12 & 0.47 & 1.0 & 0.19 & 0.42 \\ \hline
\multirow{7}{*}{\begin{tabular}[c]{@{}c@{}}EMU\\ \\ Pessimistic \\ Sensitivity\\ (20 $\mu$Jy \\ rms/beam)\end{tabular}} & \multirow{3}{*}{1 bin} & Known & 20 & 39 & 0.96 & 2.2 & 0.15 & 0.36 \\
 &  & $\Delta b_{\rm all}$ & 20 & 39 & 1.1 & 2.2 & 0.20 & 0.44 \\
 &  & $\Delta b_{\rm all}$ \& $\Delta s_{\rm all}$ & 27 & 54 & 1.1 & 2.2 & 0.20 & 0.45 \\ \cline{2-9} 
 & \multirow{4}{*}{5 bins} & Known & 1.0 & 3.0 & 0.47 & 1.1 & 0.14 & 0.35 \\
 &  & $\Delta b_{\rm all}$ & 2.3 & 5.5 & 0.59 & 1.3 & 0.19 & 0.43 \\
 &  & $\Delta b_i$ & 5.1  & 12 & 0.60 & 1.3 & 0.19 & 0.43 \\
 &  & $\Delta b_i$ \& $\Delta s_i$ & 8.4 & 19 & 0.61 & 1.4 & 0.19 & 0.44 \\ \hline
\multirow{7}{*}{\begin{tabular}[c]{@{}c@{}}EMU\\ \\ Early\\ Results\\ (100 $\mu$Jy \\ rms/beam,\\ 2000 deg$^2$)\end{tabular}} & \multirow{3}{*}{1 bin} & Known & 430 & 870 & 6.1 & 14 & 0.23 & 0.62 \\
 &  & $\Delta b_{\rm all}$ &  13000 & 27000 & 7.1 & 14 & 0.23 & 0.62 \\
 &  & $\Delta b_{\rm all}$ \& $\Delta s_{\rm all}$ & 15000 & 31000 & 9.2 & 14 & 0.23 & 0.62 \\ \cline{2-9} 
 & \multirow{4}{*}{5 bins} & Known & 6.2 & 19 & 2.9 & 7.1 & 0.23 & 0.64 \\
 &  & $\Delta b_{\rm all}$ & 61 & 140 & 3.5 & 7.8 & 0.23 & 0.64 \\
 &  & $\Delta b_i$ &88  & 200 & 3.5 & 7.9 & 0.23 & 0.64 \\
 &  & $\Delta b_i$ \& $\Delta s_i$ & 100 & 220 & 3.5 & 7.9 & 0.23 & 0.65 \\ \hline
\multirow{7}{*}{\begin{tabular}[c]{@{}c@{}}SKA-2\\ \\ (1 $\mu$Jy\\ rms/beam)\end{tabular}} & \multirow{3}{*}{1 bin} & Known & 12 & 24 & 0.79 & 1.8 & 0.15 & 0.35 \\
 &  & $\Delta b_{\rm all}$ & 46 & 91 & 0.82 & 1.8 & 0.20 & 0.44 \\
 &  & $\Delta b_{\rm all}$ \& $\Delta s_{\rm all}$ & 48 & 96 & 1.0 & 1.9 & 0.20 & 0.44 \\ \cline{2-9} 
 & \multirow{4}{*}{5 bins} & Known & 0.20 & 0.59 & 0.17 & 0.39 & 0.092 & 0.21 \\
 &  & $\Delta b_{\rm all}$ & 0.29 & 0.80 & 0.20 & 0.46 & 0.13 & 0.29 \\
 &  & $\Delta b_i$ & 1.4 & 3.2 & 0.32 & 0.73 & 0.17 & 0.39 \\
 &  & $\Delta b_i$ \& $\Delta s_i$ & 1.7 & 4.0 & 0.33 & 0.74 & 0.17 & 0.39 \\ \hline
\end{tabular}
\caption{Same as Table~\ref{tab:constr_fLN} but for $\mu_0$ and $\gamma_0$ assuming a phenomenological parameterization of modified gravity as a  $\Lambda$CDM+$\mu_0$+$\gamma_0$ model. In this case, the external data is Planck+BAO+RSD.}
\label{tab:constr_MG}
\end{table}

% \usepackage{multirow}
\begin{table}[]
\centering
\begin{tabular}{|c|c|c|cc|c|c|c|c|}
\hline
\multirow{2}{*}{Survey} & \multirow{2}{*}{\begin{tabular}[c]{@{}c@{}}\# redshift \\ bins\end{tabular}} & \multirow{2}{*}{\begin{tabular}[c]{@{}c@{}}Bias \\ uncertainty\end{tabular}} & \multicolumn{6}{c|}{\begin{tabular}[c]{@{}c@{}}Data combination and constraints on \\ extended models\end{tabular}} \\ \cline{4-9} 
 &  &  & \multicolumn{2}{c}{Galaxy Clustering (GC)} & \multicolumn{2}{c|}{GC+ISW} & \multicolumn{2}{c|}{GC+ISW+Planck$^*$} \\ \hline
\multirow{32}{*}{\begin{tabular}[c]{@{}c@{}}EMU\\ with\\ multi-tracer\\ \\ Design\\ Sensitivity\\ (10 $\mu$Jy\\  rms/beam)\\ \\ Using \\ T-RECS\end{tabular}} & \multicolumn{2}{c|}{} & \multicolumn{2}{c|}{$f_{\rm NL}$} & \multicolumn{2}{c|}{$f_{\rm NL}$} & \multicolumn{2}{c|}{$f_{\rm NL}$} \\ \cline{2-9} 
 & \multirow{3}{*}{1 bin} & Known & \multicolumn{2}{c|}{4.0} & \multicolumn{2}{c|}{3.9} & \multicolumn{2}{c|}{2.9} \\
 &  & $\Delta b_{\rm all}$ & \multicolumn{2}{c|}{4.2} & \multicolumn{2}{c|}{4.0} & \multicolumn{2}{c|}{3.0} \\
 &  & $\Delta b_{\rm all}$ \& $\Delta s_{\rm all}$ & \multicolumn{2}{c|}{5.1} & \multicolumn{2}{c|}{4.9} & \multicolumn{2}{c|}{3.2} \\ \cline{2-9} 
 & \multirow{4}{*}{5 bins} & Known & \multicolumn{2}{c|}{7.1} & \multicolumn{2}{c|}{6.6} & \multicolumn{2}{c|}{4.4} \\
 &  & $\Delta b_{\rm all}$ & \multicolumn{2}{c|}{7.3} & \multicolumn{2}{c|}{6.7} & \multicolumn{2}{c|}{4.5} \\
 &  & $\Delta b_i$ & \multicolumn{2}{c|}{7.6} & \multicolumn{2}{c|}{7.0} & \multicolumn{2}{c|}{4.6} \\
 &  & $\Delta b_i$ \& $\Delta s_i$ & \multicolumn{2}{c|}{8.1} & \multicolumn{2}{c|}{7.5} & \multicolumn{2}{c|}{4.6} \\ \cline{2-9} 
 & \multicolumn{2}{c|}{} & \multicolumn{1}{c|}{$w_0$} & $w_a$ & $w_0$ & $w_a$ & $w_0$ & $w_a$ \\ \cline{2-9} 
 & \multirow{3}{*}{1 bin} & Known & \multicolumn{1}{c|}{1.1} & 3.2 & 0.93 & 2.5 & 0.18 & 0.46 \\
 &  & $\Delta b_{\rm all}$ & \multicolumn{1}{c|}{1.2} & 4.8 & 0.98 & 0.27 & 0.19 & 0.47 \\
 &  & $\Delta b_{\rm all}$ \& $\Delta s_{\rm all}$ & \multicolumn{1}{c|}{3.0} & 11 & 1.7 & 4.5 & 0.26 & 0.71 \\ \cline{2-9} 
 & \multirow{4}{*}{5 bins} & Known & \multicolumn{1}{c|}{0.56} & 1.6 & 0.49 & 1.4 & 0.19 & 0.48 \\
 &  & $\Delta b_{\rm all}$ & \multicolumn{1}{c|}{0.57} & 1.7 & 0.50 & 1.4 & 0.19 & 0.49 \\
 &  & $\Delta b_i$ & \multicolumn{1}{c|}{1.2} & 4.4 & 0.78 & 2.1 & 0.25 & 0.68 \\
 &  & $\Delta b_i$ \& $\Delta s_i$ & \multicolumn{1}{c|}{1.7} & 5.7 & 1.2 & 2.8 & 0.25 & 0.70 \\ \cline{2-9} 
 & \multicolumn{2}{c|}{} & \multicolumn{2}{c|}{$100\times \Omega_k$} & \multicolumn{2}{c|}{$100\times \Omega_k$} & \multicolumn{2}{c|}{$100\times \Omega_k$} \\ \cline{2-9} 
 & \multirow{3}{*}{1 bin} & Known & \multicolumn{2}{c|}{3.9} & \multicolumn{2}{c|}{3.6} & \multicolumn{2}{c|}{0.16} \\
 &  & $\Delta b_{\rm all}$ & \multicolumn{2}{c|}{4.9} & \multicolumn{2}{c|}{3.9} & \multicolumn{2}{c|}{0.16} \\
 &  & $\Delta b_{\rm all}$ \& $\Delta s_{\rm all}$ & \multicolumn{2}{c|}{4.9} & \multicolumn{2}{c|}{4.1} & \multicolumn{2}{c|}{0.19} \\ \cline{2-9} 
 & \multirow{4}{*}{5 bins} & Known & \multicolumn{2}{c|}{4.1} & \multicolumn{2}{c|}{3.6} & \multicolumn{2}{c|}{0.17} \\
 &  & $\Delta b_{\rm all}$ & \multicolumn{2}{c|}{4.5} & \multicolumn{2}{c|}{3.8} & \multicolumn{2}{c|}{0.17} \\
 &  & $\Delta b_i$ & \multicolumn{2}{c|}{11} & \multicolumn{2}{c|}{6.7} & \multicolumn{2}{c|}{0.19} \\
 &  & $\Delta b_i$ \& $\Delta s_i$ & \multicolumn{2}{c|}{13} & \multicolumn{2}{c|}{7.1} & \multicolumn{2}{c|}{0.19} \\ \cline{2-9} 
 & \multicolumn{2}{c|}{} & \multicolumn{1}{c|}{$\mu_0$} & $\gamma_0$ & $\mu_0$ & $\gamma_0$ & $\mu_0$ & $\gamma_0$ \\ \cline{2-9} 
 & \multirow{3}{*}{1 bin} & Known & \multicolumn{1}{c|}{0.93} & 2.6 & 0.50 & 1.1 & 0.14 &  0.33  \\
 &  & $\Delta b_{\rm all}$ & \multicolumn{1}{c|}{3.1} &  6.4& 0.69 & 1.6 & 0.15 & 0.34 \\
 &  & $\Delta b_{\rm all}$ \& $\Delta s_{\rm all}$ & \multicolumn{1}{c|}{8.9} & 18 & 0.70 & 1.6 & 0.19 & 0.43 \\ \cline{2-9} 
 & \multirow{4}{*}{5 bins} & Known & \multicolumn{1}{c|}{0.36} & 0.90 & 0.31 & 0.71 & 0.12 & 0.29 \\
 &  & $\Delta b_{\rm all}$ & \multicolumn{1}{c|}{0.39} & 0.95 & 0.33 & 0.75 & 0.13 & 0.31 \\
 &  & $\Delta b_i$ & \multicolumn{1}{c|}{2.4} & 5.2 & 0.60 & 1.3 & 0.19 & 0.43 \\
 &  & $\Delta b_i$ \& $\Delta s_i$ & \multicolumn{1}{c|}{2.8} & 6.1 & 0.61 & 1.3 & 0.19 & 0.43 \\ \hline
\end{tabular}
\caption{Same as in Tables~\ref{tab:constr_fLN}, \ref{tab:constr_w0wa}, \ref{tab:constr_omegak} and \ref{tab:constr_MG}, but only for the EMU survey at design sensitivity and using SFGs and AGNs as different tracers. ``+Planck$^*$'' means Planck+BAO in all case but in the $\Lambda$CDM+$\mu_0$+$\gamma_0$ constraints, where means Planck+BAO+RSD.}
\label{tab:constr_multi}
\end{table}

\section{Comparison with $S^3$ simulation}\label{App:S3}
Prior to the release of the T-RECS catalogues, most of the forecast analysis for galaxy radio surveys were done using the $S^3$ simulations~\cite{Wilman_S3}. In order to ease the comparison with those studies and also to illustrate the differences between both set of simulations we repeat the study using $S^3$ for the four models under considerations and assuming the design sensitivity realization of EMU. We consider a case with all the galaxies used as a single tracer and another one using them as two different tracers. The differences between the results using each simulation also gives an estimate of the inherent uncertainty of the cosmological forecast using simulations.

\begin{figure}
\centering
\includegraphics[width=0.49\linewidth]{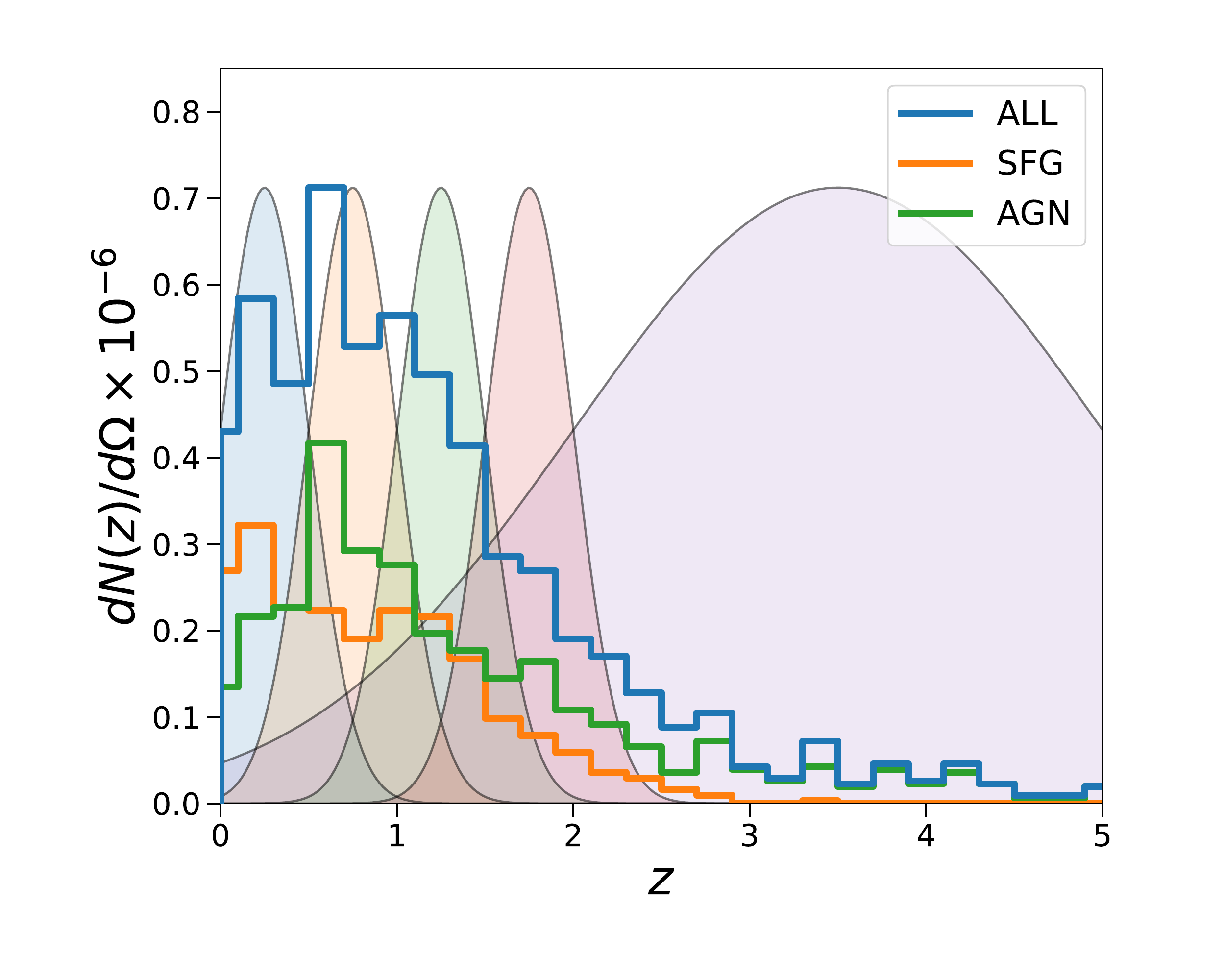}
\includegraphics[width=0.49\linewidth]{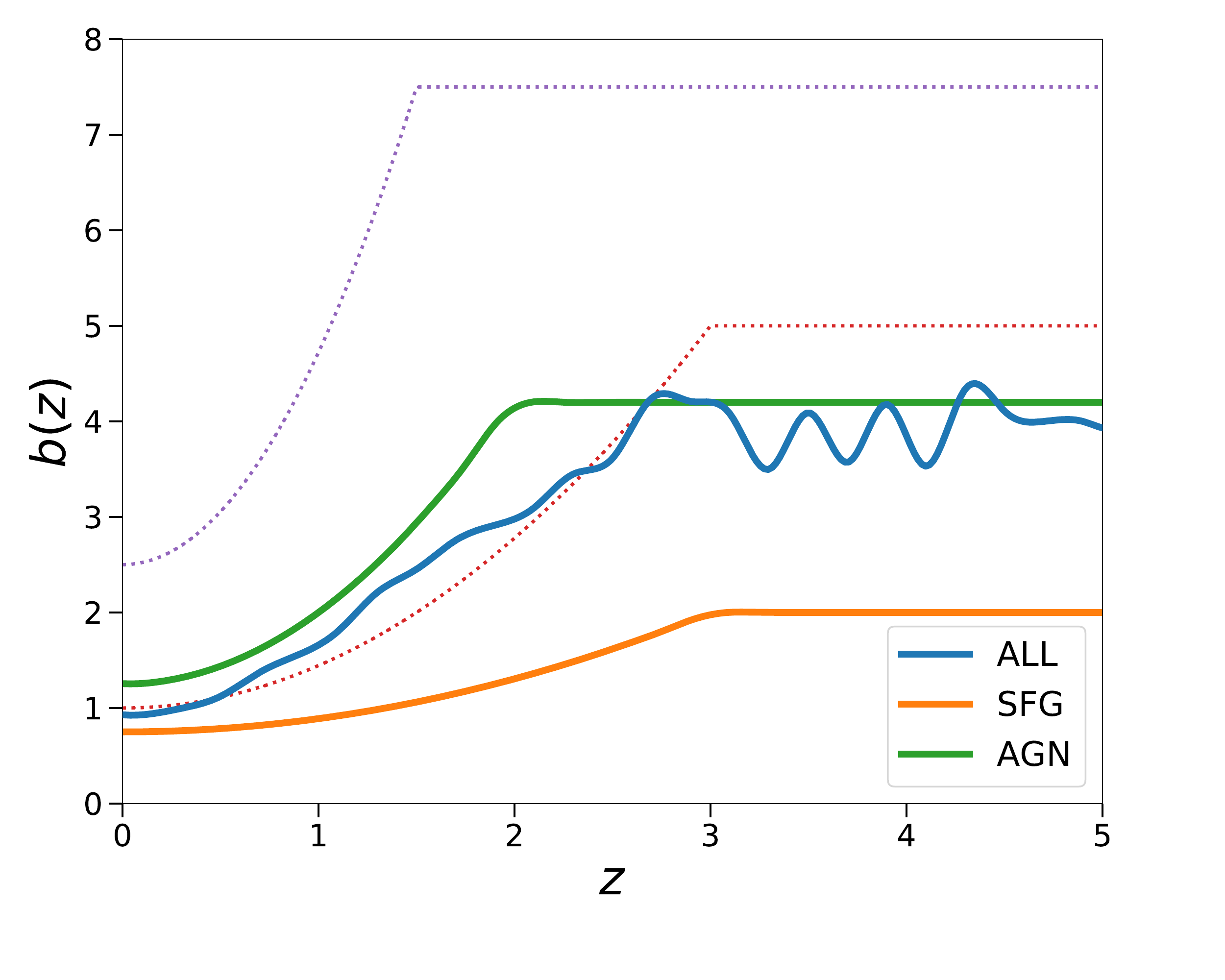}
\caption{ \textit{Left}: $dN/dzd\Omega$ for the galaxies observed assuming a rms flux/beam of 10$\mu$Jy from $S^3$ simulation, always requiring a signal of the flux of five times larger than the rms/beam to consider a detection. We also plot the Gaussian window functions of the five redshift bins considered to compute the cosmological observables.  \textit{Right} theoretical galaxy bias for SFGs and AGNs and the corresponding weighted total from $S^3$. We show the quantities related to the whole sample in blue, with SFGs alone in orange and with AGNs alone in green. We also show the galaxy bias for radio quiet quasars and Fanoroff-Riley type-II in red and purple dotted lines, respectively. }
\label{fig:specs_S3}
\end{figure}

 We show the relevant quantities obtained from $S^3$ in Figure~\ref{fig:specs_S3}. $S^3$ further subdivides the SFGs into star burst galaxies and star forming galaxies, and the AGNs into radio quiet quasars and Fanoroff-Riley type-I and type-II radio galaxies. These five populations of galaxies have been used as different tracers for studies in the literature (e.g.~\cite{Ferramacho_radioPNG}).
 However, an accurate discrimination among all five might be uncertain, so we prefer to proceed as with T-RECS and consider only two different tracers SFGs and AGNs. As star burst galaxies are far less abundant than star forming galaxies, we assume the galaxy bias model of the latter for the whole SFG population. Regarding AGNs, Fanoroff-Riley type-I galaxies abundance is negligible. Therefore, we only consider the bias model of the radio quiet quasars and Fanoroff-Riley type-II and use an approximated weighted mean. The galaxy bias  for both radio quiet quasars and Fanoroff-Riley type-II are shown in red and purple dotted lines, respectively. The quantities shown in Figure~\ref{fig:specs_S3} are to be compared with their equivalents using T-RECS (Figure~\ref{fig:specs}). 
The total number of galaxies above the threshold of detection is similar for both galaxies. However,  the distribution in $S^3$ is broader than in T-RECS, and there are also more AGNs than SFGs, especially at large redshifts. This is very important, since it affects significantly the values of the weighted mean of $b(z)$ when all galaxies are considered as a single tracer. It also reverses which tracer has more weight in the multi-tracer analysis. These differences are the responsible of the discrepancies between $S^3$ and T-RECS.
 
The results using $S^3$ are reported in Table~\ref{tab:constr_S3} and~\ref{tab:constr_S3_multi}. Since the average bias is larger in $S^3$, the signal to noise ratio of the galaxy power spectra and ISW are larger too. Therefore, the constraints on $w_0$ and $w_a$ coming only from the galaxy power spectra are $\sim 30\%$ better using $S^3$, although the difference is negligible when ISW is included in the analysis. Also the magnitude of the impact of marginalizing over the galaxy and magnification biases is similar using both simulations. The gains using two different tracers to constrain $(w_0w_a)$CDM are similar too. In the case of $\Lambda$CDM+$\mu_0$+$\gamma_0$, the differences between using $S^3$ and T-RECS are not significant.

However, since the galaxy bias has so large impact in the signal of local PNG, there is a significant difference in the constraints on $f_{\rm NL}$ using each simulation.  Using $S^3$ the constraints are $\sim 3-4$ times better using all galaxies as a single tracer and combining the galaxy power spectra and the ISW. Then, EMU alone would be able to improve current bounds on $f_{\rm NL}$, even without performing a multi-tracer analysis , measuring $\sigma_{f_{\rm NL}}\sim 5$. Using two different tracers, the difference between the results from each simulation assuming a single redshift bin is negligible. However, as considering $S^3$ AGNs are more abundant than SFGs and they do not suffer from low numbers in the high redshift bins, the constraints using five redhsift bins improve with respect to those using only one (as it was the case considering T-RECS). %\JLB{Comment MG}

% \usepackage{multirow}
\begin{table}[]
\centering
\begin{tabular}{|c|c|c|cc|c|c|c|c|}
\hline
\multirow{2}{*}{Survey} & \multirow{2}{*}{\begin{tabular}[c]{@{}c@{}}\# redshift \\ bins\end{tabular}} & \multirow{2}{*}{\begin{tabular}[c]{@{}c@{}}Bias \\ uncertainty\end{tabular}} & \multicolumn{6}{c|}{\begin{tabular}[c]{@{}c@{}}Data combination and constraints on \\ extended models\end{tabular}} \\ \cline{4-9} 
 &  &  & \multicolumn{2}{c}{Galaxy Clustering (GC)} & \multicolumn{2}{c|}{GC+ISW} & \multicolumn{2}{c|}{GC+ISW+Planck$^*$} \\ \hline
\multirow{32}{*}{\begin{tabular}[c]{@{}c@{}}EMU\\ Design\\ Sensitivity\\ (10 $\mu$Jy\\  rms/beam)\\ \\ Using $S^3$ \\ simulation\end{tabular}} & \multicolumn{2}{c|}{} & \multicolumn{2}{c|}{$f_{\rm NL}$} & \multicolumn{2}{c|}{$f_{\rm NL}$} & \multicolumn{2}{c|}{$f_{\rm NL}$} \\ \cline{2-9} 
 & \multirow{3}{*}{1 bin} & Known & \multicolumn{2}{c|}{19} & \multicolumn{2}{c|}{18} & \multicolumn{2}{c|}{5.4} \\
 &  & $\Delta b_{\rm all}$ & \multicolumn{2}{c|}{29} & \multicolumn{2}{c|}{19} & \multicolumn{2}{c|}{5.5} \\
 &  & $\Delta b_{\rm all}$ \& $\Delta s_{\rm all}$ & \multicolumn{2}{c|}{49} & \multicolumn{2}{c|}{40} & \multicolumn{2}{c|}{6.0} \\ \cline{2-9} 
 & \multirow{4}{*}{5 bins} & Known & \multicolumn{2}{c|}{5.7} & \multicolumn{2}{c|}{5.5} & \multicolumn{2}{c|}{3.6} \\
 &  & $\Delta b_{\rm all}$ & \multicolumn{2}{c|}{5.8} & \multicolumn{2}{c|}{5.6} & \multicolumn{2}{c|}{3.6} \\
 &  & $\Delta b_i$ & \multicolumn{2}{c|}{6.0} & \multicolumn{2}{c|}{5.8} & \multicolumn{2}{c|}{3.8} \\
 &  & $\Delta b_i$ \& $\Delta s_i$ & \multicolumn{2}{c|}{6.3} & \multicolumn{2}{c|}{6.1} & \multicolumn{2}{c|}{3.8} \\ \cline{2-9} 
 & \multicolumn{2}{c|}{} & \multicolumn{1}{c|}{$w_0$} & $w_a$ & $w_0$ & $w_a$ & $w_0$ & $w_a$ \\ \cline{2-9} 
 & \multirow{3}{*}{1 bin} & Known & \multicolumn{1}{c|}{6.5} & 35 & 2.0 & 6.0 & 0.19 & 0.45 \\
 &  & $\Delta b_{\rm all}$ & \multicolumn{1}{c|}{13} & 66 & 2.3 & 6.1 & 0.25 & 0.71 \\
 &  & $\Delta b_{\rm all}$ \& $\Delta s_{\rm all}$ & \multicolumn{1}{c|}{15} & 79 & 3.0 & 7.7 & 0.26 & 0.73 \\ \cline{2-9} 
 & \multirow{4}{*}{5 bins} & Known & \multicolumn{1}{c|}{0.43} & 1.3 & 0.40 & 1.1 & 0.17 & 0.43 \\
 &  & $\Delta b_{\rm all}$ & \multicolumn{1}{c|}{0.44} & 2.0 & 0.40 & 1.2 & 0.19 & 0.57 \\
 &  & $\Delta b_i$ & \multicolumn{1}{c|}{1.0} & 4.0 & 0.71 & 2.0 & 0.24 & 0.67 \\
 &  & $\Delta b_i$ \& $\Delta s_i$ & \multicolumn{1}{c|}{1.6} & 5.8 & 1.0 & 2.6 & 0.25 & 0.70 \\ \cline{2-9} 
 & \multicolumn{2}{c|}{} & \multicolumn{2}{c|}{$100\times \Omega_k$} & \multicolumn{2}{c|}{$100\times \Omega_k$} & \multicolumn{2}{c|}{$100\times \Omega_k$} \\ \cline{2-9} 
 & \multirow{3}{*}{1 bin} & Known & \multicolumn{2}{c|}{7.6} & \multicolumn{2}{c|}{7.1} & \multicolumn{2}{c|}{0.16} \\
 &  & $\Delta b_{\rm all}$ & \multicolumn{2}{c|}{7.8} & \multicolumn{2}{c|}{7.1} & \multicolumn{2}{c|}{0.19} \\
 &  & $\Delta b_{\rm all}$ \& $\Delta s_{\rm all}$ & \multicolumn{2}{c|}{7.9} & \multicolumn{2}{c|}{7.2} & \multicolumn{2}{c|}{0.19} \\ \cline{2-9} 
 & \multirow{4}{*}{5 bins} & Known & \multicolumn{2}{c|}{4.7} & \multicolumn{2}{c|}{3.5} & \multicolumn{2}{c|}{0.16} \\
 &  & $\Delta b_{\rm all}$ & \multicolumn{2}{c|}{8.3} & \multicolumn{2}{c|}{5.7} & \multicolumn{2}{c|}{0.18} \\
 &  & $\Delta b_i$ & \multicolumn{2}{c|}{9.0} & \multicolumn{2}{c|}{5.8} & \multicolumn{2}{c|}{0.19} \\
 &  & $\Delta b_i$ \& $\Delta s_i$ & \multicolumn{2}{c|}{9.9} & \multicolumn{2}{c|}{6.0} & \multicolumn{2}{c|}{0.19} \\ \cline{2-9} 
 & \multicolumn{2}{c|}{} & \multicolumn{1}{c|}{$\mu_0$} & $\gamma_0$ & $\mu_0$ & $\gamma_0$ & $\mu_0$ & $\gamma_0$ \\ \cline{2-9} 
 & \multirow{3}{*}{1 bin} & Known & \multicolumn{1}{c|}{10} & 23 & 1.1 & 2.6 & 0.15 & 0.35 \\
 &  & $\Delta b_{\rm all}$ & \multicolumn{1}{c|}{10} & 23 & 1.2 & 2.6 & 0.20 & 0.45 \\
 &  & $\Delta b_{\rm all}$ \& $\Delta s_{\rm all}$ & \multicolumn{1}{c|}{12} & 26 & 1.3 & 2.7 & 0.20 & 0.45 \\ \cline{2-9} 
 & \multirow{4}{*}{5 bins} & Known & \multicolumn{1}{c|}{0.37} & 1.1  & 0.28 & 0.67 & 0.12 & 0.28 \\
 &  & $\Delta b_{\rm all}$ & \multicolumn{1}{c|}{1.2} & 2.8 & 0.46 & 1.0 & 0.18 & 0.40 \\
 &  & $\Delta b_i$ & \multicolumn{1}{c|}{2.9} & 6.4 & 0.49 & 1.1 & 0.19 & 0.42 \\
 &  & $\Delta b_i$ \& $\Delta s_i$ & \multicolumn{1}{c|}{4.4} & 9.8 & 0.50 & 1.1 & 0.19 & 0.42 \\ \hline
\end{tabular}
\caption{Same as in Tables~\ref{tab:constr_fLN}, \ref{tab:constr_w0wa}, \ref{tab:constr_omegak} and \ref{tab:constr_MG}, but only for the EMU survey at design sensitivity and using the $S^3$ simulation~\cite{Wilman_S3} instead of T-RECS to obtain the redshift distribution of sources and the bias. ``+Planck$^*$'' means Planck+BAO in all case but in the $\Lambda$CDM+$\mu_0$+$\gamma_0$ constraints, where means Planck+BAO+RSD.} 
\label{tab:constr_S3}
\end{table}

% \usepackage{multirow}
\begin{table}[]
\centering
\begin{tabular}{|c|c|c|cc|c|c|c|c|}
\hline
\multirow{2}{*}{Survey} & \multirow{2}{*}{\begin{tabular}[c]{@{}c@{}}\# redshift \\ bins\end{tabular}} & \multirow{2}{*}{\begin{tabular}[c]{@{}c@{}}Bias \\ uncertainty\end{tabular}} & \multicolumn{6}{c|}{\begin{tabular}[c]{@{}c@{}}Data combination and constraints on \\ extended models\end{tabular}} \\ \cline{4-9} 
 &  &  & \multicolumn{2}{c}{Galaxy Clustering (GC)} & \multicolumn{2}{c|}{GC+ISW} & \multicolumn{2}{c|}{GC+ISW+Planck$^*$} \\ \hline
\multirow{32}{*}{\begin{tabular}[c]{@{}c@{}}EMU\\ with\\ multi-tracer\\ \\ Design\\ Sensitivity\\ (10 $\mu$Jy\\  rms/beam)\\ \\ Using \\ $S^3$\end{tabular}} & \multicolumn{2}{c|}{} & \multicolumn{2}{c|}{$f_{\rm NL}$} & \multicolumn{2}{c|}{$f_{\rm NL}$} & \multicolumn{2}{c|}{$f_{\rm NL}$} \\ \cline{2-9} 
 & \multirow{3}{*}{1 bin} & Known & \multicolumn{2}{c|}{4.3} & \multicolumn{2}{c|}{4.3} & \multicolumn{2}{c|}{3.0} \\
 &  & $\Delta b_{\rm all}$ & \multicolumn{2}{c|}{4.6} & \multicolumn{2}{c|}{4.3} & \multicolumn{2}{c|}{3.1} \\
 &  & $\Delta b_{\rm all}$ \& $\Delta s_{\rm all}$ & \multicolumn{2}{c|}{6.1} & \multicolumn{2}{c|}{5.7} & \multicolumn{2}{c|}{3.1} \\ \cline{2-9} 
 & \multirow{4}{*}{5 bins} & Known & \multicolumn{2}{c|}{3.8} & \multicolumn{2}{c|}{3.7} & \multicolumn{2}{c|}{3.0} \\
 &  & $\Delta b_{\rm all}$ & \multicolumn{2}{c|}{4.0} & \multicolumn{2}{c|}{3.8} & \multicolumn{2}{c|}{3.2} \\
 &  & $\Delta b_i$ & \multicolumn{2}{c|}{4.9} & \multicolumn{2}{c|}{4.6} & \multicolumn{2}{c|}{3.5} \\
 &  & $\Delta b_i$ \& $\Delta s_i$ & \multicolumn{2}{c|}{4.8} & \multicolumn{2}{c|}{3.7} & \multicolumn{2}{c|}{3.2} \\ \cline{2-9} 
 & \multicolumn{2}{c|}{} & \multicolumn{1}{c|}{$w_0$} & $w_a$ & $w_0$ & $w_a$ & $w_0$ & $w_a$ \\ \cline{2-9} 
 & \multirow{3}{*}{1 bin} & Known & \multicolumn{1}{c|}{1.2} & 3.5 & 0.85 & 2.4 & 0.17 & 0.43 \\
 &  & $\Delta b_{\rm all}$ & \multicolumn{1}{c|}{1.5} & 6.2 & 1.0 & 2.9 & 0.20 & 0.47 \\
 &  & $\Delta b_{\rm all}$ \& $\Delta s_{\rm all}$ & \multicolumn{1}{c|}{2.8} & 12 & 1.3 & 3.6 & 0.26 & 0.71 \\ \cline{2-9} 
 & \multirow{4}{*}{5 bins} & Known & \multicolumn{1}{c|}{0.40} & 1.2 & 0.35 & 1.0 & 0.16 & 0.43 \\
 &  & $\Delta b_{\rm all}$ & \multicolumn{1}{c|}{0.41} & 1.3 & 0.36 & 1.1 & 0.18 & 0.47 \\
 &  & $\Delta b_i$ & \multicolumn{1}{c|}{0.98} & 3.8 & 0.55 & 1.6 & 0.24 & 0.65 \\
 &  & $\Delta b_i$ \& $\Delta s_i$ & \multicolumn{1}{c|}{1.4} & 5.9 & 0.67 & 1.8 & 0.24 & 0.67 \\ \cline{2-9} 
 & \multicolumn{2}{c|}{} & \multicolumn{2}{c|}{$100\times \Omega_k$} & \multicolumn{2}{c|}{$100\times \Omega_k$} & \multicolumn{2}{c|}{$100\times \Omega_k$} \\ \cline{2-9} 
 & \multirow{3}{*}{1 bin} & Known & \multicolumn{2}{c|}{4.3} & \multicolumn{2}{c|}{3.7} & \multicolumn{2}{c|}{0.16} \\
 &  & $\Delta b_{\rm all}$ & \multicolumn{2}{c|}{6.0} & \multicolumn{2}{c|}{4.5} & \multicolumn{2}{c|}{0.16} \\
 &  & $\Delta b_{\rm all}$ \& $\Delta s_{\rm all}$ & \multicolumn{2}{c|}{6.1} & \multicolumn{2}{c|}{4.6} & \multicolumn{2}{c|}{0.19} \\ \cline{2-9} 
 & \multirow{4}{*}{5 bins} & Known & \multicolumn{2}{c|}{3.2} & \multicolumn{2}{c|}{2.8} & \multicolumn{2}{c|}{0.16} \\
 &  & $\Delta b_{\rm all}$ & \multicolumn{2}{c|}{3.7} & \multicolumn{2}{c|}{3.1} & \multicolumn{2}{c|}{0.17} \\
 &  & $\Delta b_i$ & \multicolumn{2}{c|}{9.6} & \multicolumn{2}{c|}{5.3} & \multicolumn{2}{c|}{0.19} \\
 &  & $\Delta b_i$ \& $\Delta s_i$ & \multicolumn{2}{c|}{11} & \multicolumn{2}{c|}{5.7} & \multicolumn{2}{c|}{0.19} \\ \cline{2-9} 
 & \multicolumn{2}{c|}{} & \multicolumn{1}{c|}{$\mu_0$} & $\gamma_0$ & $\mu_0$ & $\gamma_0$ & $\mu_0$ & $\gamma_0$ \\ \cline{2-9} 
 & \multirow{3}{*}{1 bin} & Known & \multicolumn{1}{c|}{0.74} & 1.9 & 0.45 & 1.0 & 0.14 & 0.31 \\
 &  & $\Delta b_{\rm all}$ & \multicolumn{1}{c|}{3.1} & 6.3 & 0.64 & 1.4 & 0.14 & 0.32  \\
 &  & $\Delta b_{\rm all}$ \& $\Delta s_{\rm all}$ & \multicolumn{1}{c|}{5.8} & 12 & 0.65 & 1.5 & 0.19 & 0.43 \\ \cline{2-9} 
 & \multirow{4}{*}{5 bins} & Known & \multicolumn{1}{c|}{0.29} & 0.73 & 0.24 & 0.56 & 0.12 & 0.29 \\
 &  & $\Delta b_{\rm all}$ & \multicolumn{1}{c|}{0.32} & 0.77 & 0.26 & 0.60 & 0.12 & 0.29 \\
 &  & $\Delta b_i$ & \multicolumn{1}{c|}{1.2} & 2.7  & 0.44 & 0.99 & 0.18 & 0.41 \\
 &  & $\Delta b_i$ \& $\Delta s_i$ & \multicolumn{1}{c|}{1.7} & 3.7 & 0.46 & 1.0 & 0.19 & 0.41 \\ \hline
\end{tabular}
\caption{Same as in Table~\ref{tab:constr_multi}, but using the $S^3$ simulation~\cite{Wilman_S3} instead of T-RECS to obtain the redshift distribution of sources and the bias.}
\label{tab:constr_S3_multi}
\end{table}

\end{document}